%% file: diplom_arxiv.tex
\documentclass[a4paper,11pt,reqno]{amsbook}

\usepackage[latin1]{inputenc}
\usepackage[english]{babel}
\usepackage[pdftex,bookmarks,plainpages=false,pdfpagelabels]{hyperref}
\usepackage[twoside,top=3cm,bottom=3cm,left=2.85cm,right=2.65cm,nomarginpar]{geometry}
\usepackage{graphicx}
\usepackage{amsmath,amssymb,amsthm}
\usepackage{fancyhdr}
\usepackage{footmisc}
\hypersetup{colorlinks 	= true,
			linkcolor 	= black,
			anchorcolor = black,
    		citecolor 	= blue,
    		filecolor 	= black,
    		pagecolor 	= black,
    		urlcolor 	= black}

\renewcommand{\chaptermark}[1]%
         {\markleft{#1}}

\renewcommand{\sectionmark}[1]%
         {\markright{\thechapter.\thesection\ #1}}

\numberwithin{equation}{chapter}
\numberwithin{figure}{chapter}

\begin{document}
\newcommand{\supp}[1]{\underset{#1}{supp}\ }
\newcommand{\res}[2]{\underset{#1}{Res}\left(#2\right)}
\renewcommand{\div}[1]{\mathrm{div}\left(#1\right)}
\renewcommand{\vec}[1]{\mathbf{#1}}
\renewcommand{\Re}{\mathrm{Re}\ }
\renewcommand{\Im}{\mathrm{Im}\ }
\newtheorem{mydef}{Definition}
\newtheorem{thm}{Theorem}
\newtheorem{lemm}{Lemma}
\newtheorem{cor}{Corollary}

\frontmatter

  \thispagestyle{empty}
  \vspace*{\stretch{1}}
  {\parindent0cm
   \rule{\linewidth}{.7ex}}
  \begin{flushright}
  	\vspace*{\stretch{1}}
   	\sffamily\bfseries\LARGE On the Time-Dependent Analysis of Gamow Decay\\
   	\vspace*{\stretch{1}}
   	\sffamily\bfseries\large
   	
   	\vspace*{\stretch{1}}
  \end{flushright} 
  \rule{\linewidth}{.7ex}
  \vspace*{\stretch{5}}
  \begin{center}
   \includegraphics[width=2in]{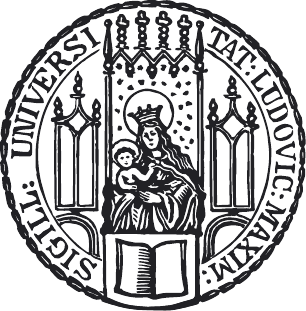}
  \end{center}
  \vspace*{\stretch{1}}
  \begin{center}
	{\sf\huge{Diplomarbeit}}\\
	\vskip .4cm
	{\sffamily\LARGE{4}}
  \end{center}
  \vspace*{1cm}
  \begin{center}{\sf\large Fakult\"at f\"ur Physik, Ludwig-Maximilians-Universit\"at M\"unchen}\end{center}
  
  \newpage
  \thispagestyle{empty}
  \cleardoublepage
  \thispagestyle{empty}
  \begin{center}
   {\sffamily\bfseries\LARGE On the Time-Dependent Analysis of Gamow Decay}\\
	\vskip .5cm
	{\large\sc \footnote{{\sc Mathematisches Institut, LMU M\"unchen,} eMail: grummt@math.lmu.de}}\\
	\vskip .2cm
	{June 10, 2009}
	\vskip 1cm
  \end{center}
  \input{module0.tex}

  \newpage
  \thispagestyle{empty}
  \vspace*{\stretch{1}}
  \begin{flushleft}
  \end{flushleft}
  \cleardoublepage
VC
      {Finsterwalde}                   	
      {Fakult\"at f\"ur Physik}		    
      {M\"unchen 2009}                 	
      {10. Juni 2009}                   
      {Prof. Dr. D. D\"urr}            	
      {Prof. Dr. P. Mayr}              	

\tableofcontents
\cleardoublepage

\mainmatter
\setcounter{page}{1}

\input{module1.tex}\newpage
\input{module2.tex}\newpage
\input{module32.tex}\newpage
\input{module33.tex}\newpage
\input{module4.tex}

\backmatter
\bibliographystyle{plain}
\bibliography{citations}







\end{document}

%% file: module0.tex



\begin{center}\textbf{Abstract}\end{center}
\begin{center}
\begin{minipage}{0.8\textwidth}
Gamow's approach to exponential decay of meta-stable particles via complex
'eigenvalues' (resonances) of a Hamiltonian is scrutinized. We explain the
sense in which the non-square-integrable 'eigenfunctions' that belong to these
resonances (Gamow functions) are relevant for the time-evolution of
square-integrable wave functions. For concreteness we study a one dimensional
square-well potential with a trapping region $K$ and the evolution of wave
functions, whose support is initially inside of $K.$ It is shown that the sum
over the first few time-evolved Gamow functions restricted to $K$ yields an
approximation for the evolution of these initial wave functions within the
trapping region. The approximation is good for all times for which exponential
decay prevails.
\vskip .4cm
{\bf Supervisor:} Prof. Dr. Detlef D\"urr
\end{minipage}
\end{center}


%% file: module1.tex




\chapter{Introduction}

The exponential law appearing in connection with the phenomenon of
$\alpha$-decay is experimentally well verified. Its quantum mechanical
description has a long-standing history and remains the subject of research in
mathematical physics to this day. George Gamow~\cite{Gamow} was the first to
study $\alpha$-decay within non-relativistic Quantum Mechanics in 1928.
He considered solutions $G$ to the stationary Schrödinger equation, which
correspond to a complex 'eigenvalue' $\epsilon=E-i\frac{\Gamma}{2}$ rather than
a real one. The Schrödinger evolution $G\,e^{-i\epsilon t}$ then immediately
implies the exponential decay of the 'Gamow function' $G$ with decay rate
$\Gamma.$ But due to the continuity equation this behavior is only possible when $G$
increases exponentially with increasing $x.$ Therefore, the Gamow function is
not square-integrable and has no direct physical relevance. This raises the
question whether there is a way to make sense of the Gamow functions.
The following argument outlines our approach to the solution of this problem.

The Schrödinger equation is a wave equation. Therefore, potentials whose core is
separated from the outside by high barriers will confine wave packets for a long
time. This is due to reflection of waves at barriers. Mathematically, this
property of the potential becomes manifest in the existence of 'almost bound
states' $G$ or, as we have previously called them, Gamow functions. We
already know that they solve
\begin{equation}\label{eq:Intro_Schroedinger}
  (H_0+V)\psi=\epsilon\psi
\end{equation}
with $\epsilon=E-i\frac{\Gamma}{2},$ where the imaginary part will be small
due to the long confinement measured by the lifetime $\Gamma^{-1}.$ 

On the other hand, the time evolution of a wave function $\psi_0$ within a
potential $V,$ which does not admit any bound states, can be determined by
expanding $\psi_0$ in so-called generalized eigenfunctions. This yields
\begin{equation}\label{eq:Intro_GFT}
  e^{-iHt}\psi_0(x)=\int \hat\psi_0(k)\phi(k,x)e^{-itk^2/2}\,dk,
\end{equation}
where the eigenfunctions $\phi(k,x)$ are bounded solutions of the stationary
Schrödinger equation~\eqref{eq:Intro_Schroedinger} for real $\epsilon=k^2/2\geq0.$
But if the 'eigenvalue' of the Gamow function is very close to the real axis,
the generalized eigenfunction $\phi(\sqrt{2E},x)$ should mimic the shape of
$G(x),$ since both functions solve equation~\eqref{eq:Intro_Schroedinger}. This
argument motivates that
\begin{equation}\label{eq:Intro_Residue}
  \phi(k,x)\sim\eta(k)G(x)
\end{equation}
for all $k$ in the vicinity of $\sqrt{2E}.$ 

Clearly, not every initial wave function $\psi_0$ will decay exponentially
within a certain time regime. But Gamow's approach suggests that a
square-integrable approximation of $G$ yields the exponential law at least
within a certain time interval. And a rough approximation is obtained by
truncating all of $G$ that lies beyond the core $K$ of the potential, which
yields
\begin{equation*}
  \hat\psi_0(k)=\int\chi_{_K}G(x)\overline{\phi(k,x)}\,dx\sim c\,\bar\eta(k).
\end{equation*}
Inserting this and relation~\eqref{eq:Intro_Residue} into the
integral~\eqref{eq:Intro_GFT}, the time evolution of the truncated Gamow
function reduces to the Fourier transform of $|\eta(k)|^2.$ Therefore, a
considerable part of this work is devoted to the properties of the
function~$\eta.$ It will turn out that, in the vicinity of~$\sqrt{2E},$ its
squared modulus has Breit-Wigner shape
\begin{equation*}
  |\eta(k)|^2=\frac{|a_{-1}|^2}{(k-\Re\sqrt{2\epsilon})^2+(\Im\sqrt{2\epsilon})^2}.
\end{equation*}
Since the Fourier transform of the Breit-Wigner distribution is the exponential
function, the time evolution of these initial wave functions is then
approximately given by
\begin{equation}\label{eq:Intro_Result}
  e^{-iHt}\psi_0(x)\sim c\,G(x)\int|\eta(k)|^2e^{-itk^2/2}\,dk\sim c\,G(x)e^{-\frac{\Gamma}{2}\,|t|}.
\end{equation}
Thereby, we have not only found an initial wave function which decays
exponentially within a certain time interval, but it also takes the shape of the
Gamow function $G.$

Now, this heuristic discussion needs to be turned into a rigorous one. In this
regard, the first major step will be taken in Chapter~\ref{ch:Eigenfunctions},
where the precise relation between the generalized eigenfunctions~$\phi$ and the
Gamow function~$G$ will be determined and where~$\eta$ will be calculated. This
will primarily rely upon the complex continuation of~$\phi$ in the variable $k.$
And the second major step will be taken in Chapter~\ref{ch:Results}, where we
will understand which initial wave functions~$\psi_0$ decay exponentially within
a certain time regime. The basic technique utilized in the proofs of this
chapter is the calculus of residues.

Regarding the question whether Gamow functions have physical relevance,
substantial progress was made by Skibsted with~\cite{Skibsted86}
and~\cite{Skibsted89}. He also showed that the truncated Gamow
function~$\chi_{_K}G$ decays exponentially within a certain time regime. And
although expansions in eigenfunctions are used in his approach as well, his
proofs do not rely on complex continuation. Nevertheless, he mentions this
method as a possible alternative~\cite[p.~593]{Skibsted86}, even though
Skibsted~\cite[p.~47]{Skibsted89} considers complex continuation as difficult if
rigorous results shall be obtained. However, the author believes that the
expansion of~$e^{-iHt}\psi_0$ in eigenfunctions together with the application of
the Residue Theorem to the resulting integral, is the most direct approach
to~$\alpha$-decay. It clearly demonstrates how exponential decay arises and
which role Gamow functions play in this regard. And in spite of the above
concerns, we will obtain rigorous results using this method, at least for the
specific potential studied in this work. In a sense, this thesis thereby
explains the necessity of what was achieved by Skibsted~\cite{Skibsted86}.
Moreover, it will complement Skibsted's work by explaining why truncated
Gamow functions are a particularly good choice for proving the exponential decay
of~$e^{-iHt}\psi_0$ within a certain time regime. 

Another question that arises in connection with Gamow's article is, in which
sense a self-adjoint operator admits complex 'eigenvalues' or, as they are
customarily called, resonances. In this respect, different definitions of
resonances have been studied. The most popular ones are reviewed by
Simon~\cite{Simon}. A recent review of Zworski~\cite{Zworski} takes the widely
accepted definition of resonances, as a pole of the resolvent $(z-H)^{-1}$ for
granted. This allows for the study of asymptotic properties of, what could be
called, the extended spectrum and the dependence of the resonances on the
geometry of configuration space.

However, the main concern of this thesis is to understand the physical relevance of
Gamow functions rather than questions regarding the extended spectrum.
Therefore, a precise definition of Gamow functions will be provided in the next
chapter. This will allow us to study their relation to physically
relevant~$L^2$-functions in the framework of a one dimensional square-well
potential. The motivation for choosing this potential was the
article~\cite{Roderich} of Garrido, Goldstein, Lukkarinen and Tumulka, whose
main objective - contrary to this thesis - is to show that an initially
localized wave packet leaves the square-well much slower than a classical
particle. But before this specific potential is introduced and used to explain
why an eigenfunction expansion serves to express~$e^{-iHt}\psi_0$ as an
integral, we will show that square-integrable wave functions generically undergo
exponential decay only within a certain time interval. In
Chapter~\ref{ch:Eigenfunctions} we will then establish the precise relation
between the generalized eigenfunctions and the Gamow functions for our specific
potential. The function~$\eta(k),$ which is calculated in this process, is of
vital importance for Chapter~\ref{ch:Results}. With its help it will be shown
that not only truncated Gamow functions, but also bound states obtained by
restricting the original Hamiltonian, decay exponentially within a certain time
interval. The bound states will then allow us to extend these results to a
general class of initial wave functions. And in this process we will understand
why truncated Gamow functions are distinguished initial wave functions. The last
chapter will then conclude with an outlook pointing to questions, which remain
open.

\subsection*{Notation and Units}

The real part of a complex number~$z$ will be denoted by~$z',$ its imaginary
part by~$z''$ and its complex conjugate by~$\bar z.$ Although, the prime will
denote the derivative with respect to a real variable from time to time, it will
always be self-evident which meaning of the prime is referred to in the
particular context. Moreover,~$O(\lambda^n)$ denotes a function~$f(\lambda)$
which is asymptotically of the same order as~$\lambda^n,$ that is
\begin{equation*}
  \limsup_{\lambda\rightarrow\infty}\left|\frac{f(\lambda)}{\lambda^n}\right|<\infty.
\end{equation*}
Apart from this, we choose units in which $\hbar=m=1.$



%% file: module2.tex



\chapter{Problem and Tools}
\label{ch:Preliminaries}

\section{Statement of the Problem}
\label{sec:Mission}

To approach the question whether the Gamow functions have physical relevance,
we will now define them and rephrase the problem in mathematically precise
terms.

Gamow derived the exponential decay law from non-relativistic Quantum Theory,
by considering functions $\psi$ that solve the stationary Schrödinger equation
\begin{equation}\label{eq:Schroedinger}
  -\frac{1}{2}\frac{d^2}{dx^2}\psi(x)+V(x)\psi(x)=\epsilon\,\psi(x)
\end{equation}
for complex rather than real $\epsilon.$ Since such a $\psi$ yields the solution
\begin{equation}\label{eq:evolution}
  \psi(t,\cdot)=e^{-i\epsilon t}\psi(\cdot)
\end{equation}
to the time dependent Schrödinger equation, it decays exponentially in time
when $\Im\epsilon$ is negative. But due to the fact that~\eqref{eq:evolution}
needs to respect the continuity equation
\begin{equation*}
  \partial_t|\psi|^2+\div{\Im\bar\psi\nabla\psi}=0
\end{equation*}
as well, the flux associated to this solution must be outgoing such that both
terms on the left hand side cancel. This motivates the following definition of
Gamow functions.

\begin{mydef}\label{def:Gamow}
  Given the one-dimensional stationary Schrödinger equation~\eqref{eq:Schroedinger}
  with a potential~$V$ having compact support and non-zero complex $\epsilon$
  having negative imaginary part. A function~$G\in\mathcal C^1(\mathbb R)$ is
  called Gamow function, if it solves~\eqref{eq:Schroedinger} and if there is a
  non-zero constant~$C$ such that
  \begin{equation*}
	G(x)=Ce^{+iz|x|}\qquad\forall\,x\in(\text{supp }V)^c,
  \end{equation*}
  where $\epsilon=\frac{1}{2}z^2=E-i\frac{\Gamma}{2}.$
\end{mydef}

Thus, Gamow functions are defined by regarding the Schrödinger equation as
ordinary differential equation on $\mathcal C^1$ rather than as an operator
identity on a Hilbert space. Since the notion of self-adjointness thereby loses
its meaning, it should not be disturbing that the Gamow~'eigenvalue'~$\epsilon$
is complex.

From Definition~\ref{def:Gamow} it becomes evident that Gamow functions do not
lie in the Hilbert space~$L^2(\mathbb R).$ For, the negative imaginary
part of $z$ directly implies that $G(x)$ must increase exponentially as $|x|$
tends to $\infty.$ Therefore, the question whether the Gamow functions have
physical relevance can be rephrased to whether there is a relation between
Gamow functions and Hilbert space solutions of the time dependent Schrödinger
equation.

As already outlined in the introduction, we will find a relation between $G$ and
the generalized eigenfunctions $\phi.$ And although these eigenfunctions do not
belong to $L^2(\mathbb R)$ either, they can be used to express the time
evolution of a square-integrable wave function $\psi_0$ as the integral
\begin{equation*}
  e^{-iHt}\psi_0=\int\hat\psi_0(k)\phi(k,\cdot)\,dk.
\end{equation*}
And this will allow us to establish a relation between physical relevant Hilbert
space solutions of the time dependent Schrödinger equation and Gamow functions.

\section{Properties of the Schrödinger Evolution}
\label{sec:Barry}

In the current section we will see that~$e^{-iHt}\psi$ can not decay
exponentially for~$t\gg1$ as well as~$t\ll1.$ This will follow from two
well-known properties of the survival probability, which is defined as
\begin{equation*}
  P_\psi(t)=|\left<\psi,e^{-iHt}\psi\right>|^2.
\end{equation*}
The proof presented here, is based upon the arguments given by Simon
in~\cite{Simon}. And it relies heavily on the spectral theorem for unbounded
self-adjoint operators (see e.g.~\cite[Chapter~2]{Davies}), such as the
Schrödinger operator. Therefore, we will briefly review the contents of this
theorem.

Casually speaking, the spectral theorem generalizes the fact that every
hermitian matrix can be diagonalized, to operators on Hilbert spaces. To make
this precise, let $H$ denote the Schrödinger operator, $\mathcal H$ denote the
Hilbert space on which it is defined and let $\sigma(H)$ denote its spectrum.
Then the spectral theorem guarantees the existence of a finite countably
additive measure~$\mu$ on~$\sigma(H)\times\mathbb N$ and the existence of a
unitary map
\begin{equation*}
  \mathcal F:\mathcal H\rightarrow L^2(\sigma(H)\times\mathbb N,\mu).
\end{equation*}
This map is such that $\mathcal FH\mathcal F^{-1}=h,$ where $h$ is the
multiplication operator associated to the function $h(n,s)=s$ with $s$ being an
element of $\sigma(H).$ Since $h$ is the analogue of a diagonal matrix, this
justifies the casual statement of the spectral theorem given above.
However,~$\mathcal F$ develops its full power only when it is applied to the
operator $f(H)$ with $f$ being an arbitrary bounded measurable function. In this
case we find analogously to the identity from above~$\mathcal Ff(H)\mathcal
F^{-1}=f(h),$ which will be used extensively in the following two proofs. The
fact that~$\mathcal F$ maps onto a direct sum of
several~$L^2(\sigma(H),\mu)$-spaces is due to the devision of~$\mathcal H$ into
cyclic subspaces, which is necessary to prove the spectral theorem. But on every
cyclic subspace,~$H$ is unitarily equivalent to a multiplication with $h(s).$
Therefore, we will neglect this detail in the sequel without causing any harm.

Having reviewed the spectral theorem, we can now turn to the first of the
above-mentioned properties of the survival probability $P_\psi.$ It concerns
the short time behavior of $e^{-iHt}\psi.$

\begin{lemm}\label{lem:Short_Time}
  Let $H$ be a self-adjoint operator and let $\psi$ be an element of its domain.
  Then~$P_\psi(t)$ is differentiable everywhere and $\dot P_\psi(0)=0.$
\end{lemm}

\begin{proof}
  The differentiability of $P_\psi(t)$ follows from the differentiability of
  \begin{equation*}
    A(t)=\left<\psi,e^{-iHt}\psi\right>.
  \end{equation*}

  Since $H$ is self-adjoint, the spectral theorem is applicable. This directly
  implies
  \begin{align}\label{eq:Amplitude}
	A(t)=\int_{\sigma(H)}|\mathcal F\psi(\epsilon)|^2e^{-i\epsilon t}\,\mu(d\epsilon).
  \end{align}
  Using dominated convergence and the inequality of Cauchy-Schwarz, we thereby see that
  $A$ is differentiable if
  \begin{equation*}
	\bigg|\int_{\sigma(H)}\epsilon|\mathcal F\psi(\epsilon)|^2e^{-i\epsilon
	t}\,\mu(d\epsilon)\bigg|=\left|\left<\psi,e^{-iHt}H\psi\right>\right|\leq\|\psi\|\,\|H\psi\|<\infty.
  \end{equation*}
  Since this is the case for all $\psi$ in the domain of $H,$ the survival
  probability $P_\psi(t)$ is differentiable for all $t$ in $\mathbb R.$

  Using the spectral theorem again, $P_\psi$ can be rewritten as
  \begin{equation*}
	|\left<\psi,e^{-iHt}\psi\right>|^2=\left<\psi,e^{-iHt}\psi\right>\left<\psi,e^{iHt}\psi\right>,
  \end{equation*}
  which implies that $P_\psi(t)=P_\psi(-t)$ for all $t$ in $\mathbb R$
  and~$\psi$ in the domain of~$H.$ Moreover, an application of the inequality
  of Cauchy-Schwarz proves that $P_\psi(t)\leq P_\psi(0).$ From these facts
  and the differentiability of $P_\psi,$ we can conclude that $\dot
  P_\psi(0)=0.$
\end{proof}

The differentiability of $P_\psi$ and the vanishing derivative at zero,
immediately imply that
\begin{equation*}
  P_\psi(t)\neq Ce^{-\Gamma t}
\end{equation*}
for small $t.$ So even if $e^{-iHt}\psi$ contains a contribution of the form
$e^{-\Gamma t}\tilde\psi,$ there must be an additive remainder that accounts
for the non-exponential behavior of $P_\psi$ for small $t.$ Apart from this
restriction on the short time behavior of $e^{-iHt}\psi,$ there is another
property of the survival probability having implications for the long time
behavior.

\begin{lemm}\label{lem:Paley}
  Let $H$ be a self-adjoint operator, whose spectrum is bounded from below.
  Then the only element $\psi$ in the domain of $H$ that satisfies
  \begin{equation*}
	P_\psi(t)\leq Ce^{-\Gamma|t|}
  \end{equation*}
  for some $C,\Gamma>0$ is the zero element.
\end{lemm}

\begin{proof}
  Since $H$ is self-adjoint and $\psi$ is an element of its domain,
  equation~\eqref{eq:Amplitude} can be utilized again such that
  \begin{align*}
	A(t)=\int_{\sigma(H)}|\mathcal F\psi(\epsilon)|^2e^{-i\epsilon
	t}\,\mu(d\epsilon)=\int_{\mathbb R}g(\epsilon)e^{-i\epsilon t}\,\mu(d\epsilon),
  \end{align*}
  where $g$ is equal to $|\mathcal F\psi|^2$ on the spectrum $\sigma(H)$ and zero
  otherwise. Therefore, $A$ is the Fourier transform of a $L^1$-function $g,$
  which is supported only on the spectrum of $H.$ By inverting the
  Fourier transform, we find
  \begin{equation*}
	g(\epsilon)=\int_{\mathbb R}A(t)e^{i\epsilon t}\,\mu(dt).
  \end{equation*}
  And if we assume that
  \begin{equation*}
	|A(t)|\leq Ce^{-\frac{\Gamma}{2}|t|}
  \end{equation*}
  for some non-zero positive constants $C$ and $\Gamma,$ this formula together
  with the theorem of dominated convergence implies that $g$ is analytic on the
  strip $\{\epsilon_0+i\eta|\epsilon_0\in\mathbb R,\eta\in(-\Gamma/2,\Gamma/2)\}.$
  But the spectrum of $H$
  is bounded from below. So $g(\epsilon)$ vanishes for all real $\epsilon$ smaller than
  the infimum of~$\sigma(H).$ Therefore, $g$ must vanish on the whole strip,
  which is only possible if~$\psi$ equals zero.
\end{proof}

That there is no exponential bound on $P_\psi$, can only be due to the
asymptotics of the survival probability and thereby due to the asymptotics of
$e^{-iHt}\psi.$ It reflects the fact that the wave function ultimately decays
polynomially instead of exponentially. This is not surprising, since it is in
accord with the following heuristic scattering theoretical argument~(for a
detailed discussion see~\cite[Remark~15.2.4]{Duerr}). The time evolved wave
function at the fixed position $x$ is given by
\begin{equation*}
  e^{-iHt}\psi(x)=t^{-1/2}\int\mathcal F\psi\left(\frac{\kappa}{\sqrt t}\right)\phi\left(\frac{\kappa}{\sqrt t},x\right)e^{-i\frac{\kappa^2}{2}}\,d\kappa,
\end{equation*}
where $\kappa/\sqrt t$ was substituted for $k.$ In the limit of $t$ tending to
$\infty$ the right hand side multiplied by $t^{1/2}$ turns out to be constant,
which can be seen heuristically by interchanging limit and integration. For
$t\gg1$ we therefore expect that
\begin{equation*}
  e^{-iHt}\psi(x)\sim c\,\mathcal F\psi(0)\phi(0,x)\,t^{-1/2}.
\end{equation*}

The previous lemmas also show that the heuristic discussion in the introduction,
which lead us to
\begin{equation*}
  e^{-iHt}\psi(x)\sim G(x)\,e^{-\frac{\Gamma}{2}|t|},
\end{equation*}
can only be valid on intermediate time scales. However, for very small as well
as very large times the error terms, hidden behind the $\sim$~symbol, will not
be negligible anymore. In Chapter~\ref{ch:Results} this heuristic statement will
be turned into a precise one.

Let us conclude this section with the following aside. If exponential decay can
prevail on intermediate time scales, we could just choose $t_0$ big enough and
consider
\begin{equation*}
  P_\psi(t_0+t)=|\langle\psi,e^{-iHt}e^{-iHt_0}\psi\rangle|^2
\end{equation*}
in order to ensure exponential decay even for times $t\ll1.$ This suggests that
by choosing the initial time $t_0$ advantageous the result of
Lemma~\ref{lem:Short_Time} can be somehow circumvented. But in fact resetting
the initial time results in the quantity
\begin{equation*}
  |\langle e^{-iHt_0}\psi,e^{-iHt}e^{-iHt_0}\psi\rangle|^2
\end{equation*}
as opposed to~$P_\psi(t_0+t),$ since we really want to compare the time evolved
wave function~$e^{-iHt}\psi_{t_0}$ to the initial wave function~$\psi_{t_0}.$
And this quantity is identical to~$P_\psi(t),$ which is why it can not decay
exponentially for times~$t\ll1$ either.

\section{Setting and Tools}
\label{sec:Tools}

\begin{figure}[ht]
  \centering
  \includegraphics[scale=0.5]{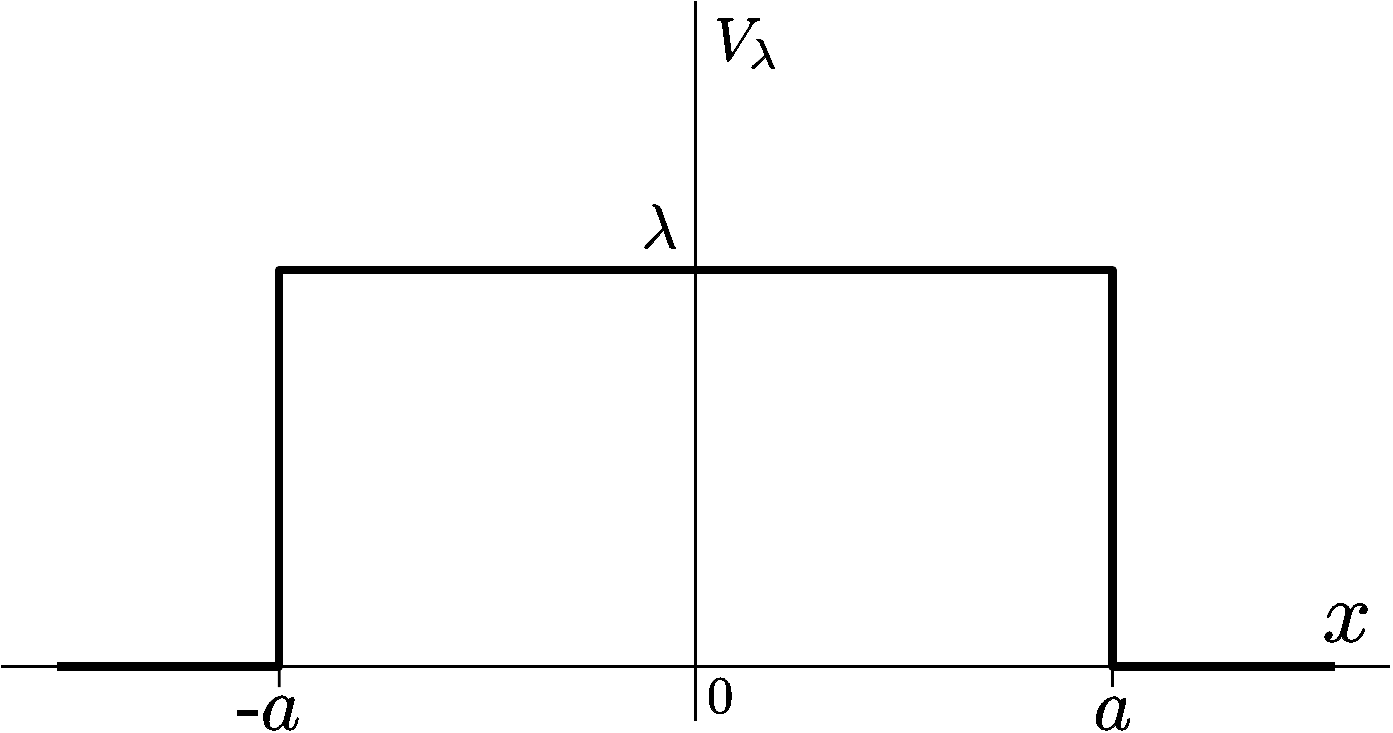}
  \caption{Plot of the potential $V_\lambda.$}
  \label{fig:Potential}
\end{figure}

The specific potential that provides the framework for this thesis is given by
\begin{equation}\label{eq:Potential}
  V_\lambda(x)=\lambda\chi_{[-a,a]}(x)\quad\text{with}\quad a,\lambda>0
\end{equation}
and is illustrated in Figure~\ref{fig:Potential}. Due to the reflection of waves
at barriers, it confines initially localized wave packets for a long time
if~$\lambda\gg1.$ As was already mentioned in the introduction,~$V_\lambda$
will therefore admit resonances. Since it is explicitly solvable
too,~$V_\lambda$ is a good choice for approaching the question whether Gamow
functions have physical relevance.

Clearly, the function $V_\lambda(x)$ defines a multiplication operator
$V_\lambda$ acting on $L^2(\mathbb R).$ It is not difficult to see that this
operator is symmetric and bounded. And in particular, $V_\lambda$ has relative
bound zero with respect to the free Schrödinger operator $H_0$ acting on
$L^2(\mathbb R).$ The Kato-Rellich theorem~(see~\cite[Theorem~1.4.2]{Davies})
therefore immediately implies that~$H_\lambda=H_0+V_\lambda$ is self-adjoint
with domain~$H^2(\mathbb R)=\mathcal D(H_0).$ And due to the self-adjointness,
the spectral theorem guarantees the existence of a map~$\mathcal F$ yielding the
spectral representation of~$H_\lambda.$ In the previous section we already used
this map, in order to diagonalize $e^{-iHt}.$ Regarding the time evolution of
the wave function~$\psi_0,$ this only yields
\begin{equation*}
  e^{-iH_\lambda t}\psi_0=\mathcal F^{-1}e^{-i(\cdot)t}\mathcal F\psi_0.
\end{equation*}
In order to get any further we need to determine $\mathcal F$ explicitly.
Fortunately, for our simple potential this can be done. One possibility
to achieve $\mathcal F,$ is to determine the integral kernel of the
resolvent~$(z-H_\lambda)^{-1}$ which rather directly yields the spectral
measure~$dP_\epsilon.$ This strategy is pursued in~\cite[Chapter~23]{Weidmann}
to which we refer for details. However, after doing the calculations one will
find that the map $\mathcal F$ yielding the spectral representation of
$H_\lambda$ can be defined as an integral operator
\begin{equation}\label{eq:GeneralizedFT}
  \mathcal F\psi(k)=\int_{\mathbb R}\psi(x)\overline{\phi(k,x)}dx
\end{equation}
with respect to the Lebesgue measure. And the integral kernel turns out to be an
element of~$L^\infty,$ which solves
\begin{equation}\label{eq:Schroedinger_GEF}
  -\frac{1}{2}\frac{d^2}{dx^2}\phi(k,x)+V\phi(k,x)=\epsilon\,\phi(k,x)\quad\text{for}\quad\epsilon=\frac{k^2}{2}\geq0
\end{equation}
and is therefore called generalized eigenfunction. Although, bona fide
eigenfunctions usually enter into the explicit expression of $\mathcal F$ as
well, this is not the case for $V_\lambda.$ For this particular potential the
pure point spectrum is empty. Thus, there are no bona fide eigenfunctions
and the spectrum is given by
\begin{equation*}
  \sigma(H_\lambda)=[0,\infty).
\end{equation*}
Before we continue, notice the similarity of equation~\eqref{eq:GeneralizedFT}
with the Fourier transform. It has earned $\mathcal F$ the name generalized
Fourier transform. Clearly, the generalized eigenfunctions $\phi$ define the
spectral representation of the perturbed Hamiltonian $H_\lambda$ in complete
analogy to the way the plane waves $e^{ikx}$ define the spectral
representation of $H_0.$ And it is often helpful to remember this analogy.

\section{Resonances and Gamow Functions}
\label{sec:Gamow}

In the present section we will determine the Gamow functions that belong to the
potential~$V_\lambda.$ Thereby, we will not only have an explicit example for
their generic definition. But in the process of calculating the Gamow functions,
we will also determine the approximate location of the corresponding
'eigenvalues'. Although these eigenvalues will be called resonances from now on,
it will only be seen in the next chapter that their defining equation coincides
with the formula which defines the poles of the $S$-Matrix thereby justifying
this name.

An ansatz for the solution of equation~\eqref{eq:Schroedinger} with potential
$V_\lambda,$ which satisfies the boundary conditions of
Definition~\ref{def:Gamow} is given by
\begin{align*}
  &G(x) = 
  \begin{cases}
	Ae^{-izx} \quad\quad&, x < -a\\
	Be^{i\tilde zx} + Ce^{-i\tilde zx} &, |x| \leq a\\
	De^{izx} &, x > a
  \end{cases}\\
  &\text{with}\quad\tilde z=\sqrt{z^2-2\lambda}.
\end{align*}
Whereas the complex square root is defined such that $\Re\sqrt w>0$ for all $w$
in $\mathbb C\setminus(-\infty,0].$ So whenever $\Im w\lessgtr0$ we know that
$\Im\sqrt w\lessgtr0.$ For $G$ to be an element of $\mathcal C^1$ as required
from the definition of Gamow functions, its first derivative and itself have to
be continuous at~$x$ equal to~$\pm a$. This will be the case when the
coefficients satisfy the following homogeneous system of linear equations
\begin{align}\label{eq:Gamow_LGS}
\begin{pmatrix}
  e^{iza} & -e^{-i\tilde za} & -e^{i\tilde za} & 0\\
  -ize^{iza} & -i\tilde ze^{-i\tilde za} & i\tilde ze^{i\tilde za} & 0\\
  0 & -e^{i\tilde za} & -e^{-i\tilde za} & e^{iza}\\
  0 & -i\tilde ze^{i\tilde za} & i\tilde ze^{-i\tilde za} & ize^{iza}
\end{pmatrix}
\begin{pmatrix}
A\\B\\C\\D
\end{pmatrix}
=0.
\end{align}
Due to the fact that this system has non-trivial solutions only when the
determinant of the matrix vanishes, the equation
\begin{equation}\label{eq:Gamow_Condition_temp}
  e^{i4a\tilde z} = \left(\frac{z+\tilde z}{z-\tilde z}\right)^2
\end{equation}
needs to have solutions for Gamow functions to exist.

\begin{lemm}\label{lem:Resonances}
  Given the potential $V_\lambda$ with fixed width $a>0,$ there is a Gamow
  function corresponding to every integer $n$ which satisfies
  \begin{equation}\label{eq:Gamow_Condition}
	1\leq n<\frac{2}{\pi}\sqrt{2a^2\lambda-1}.
  \end{equation}
  Moreover, if such an $n$ is fixed the corresponding resonance is given by
  \begin{align*}
	&z_n=\left(2\lambda+\frac{n^2\pi^2}{4a^2}\right)^{1/2}-i\frac{n^2\pi^2}{4a^3\sqrt{2\lambda}}\left(2\lambda+\frac{n^2\pi^2}{4a^2}\right)^{-1/2}+O(\lambda^{-3/2})\quad\text{and}\\
	&\tilde z_n=\frac{n\pi}{2a}-i\frac{n\pi}{2a^2\sqrt{2\lambda}}+O(\lambda^{-1})
  \end{align*}
  respectively.
\end{lemm}

\begin{proof}
  Due to the fact that $\exp(-in2\pi)=1$ for all $n$ in $\mathbb Z,$ the
  identity~\eqref{eq:Gamow_Condition_temp} represents a whole family of equations
  numbered by the integer $n.$ Using the definition of $\tilde z,$ this family
  can be expressed as
  \begin{align*}
	e^{i4a\sqrt{2\lambda}\tilde\kappa-in2\pi}=\left(\tilde\kappa+\sqrt{\tilde\kappa^2+1}\right)^4,
  \end{align*}
  where $\tilde\kappa$ denotes $\tilde z/\sqrt{2\lambda}.$ And with the help of
  some simple manipulations we can even bring it into the fixed-point form
  \begin{equation*}
	F_n(\tilde\kappa):=n\frac{\pi}{2a\sqrt{2\lambda}}-i\frac{1}{a\sqrt{2\lambda}}\ln(\tilde\kappa+\sqrt{\tilde\kappa^2+1})=\tilde\kappa,
  \end{equation*}
  where $\ln$ denotes the principal branch of the complex logarithm. Therefore,
  the Banach Fixed Point Theorem (see~\cite[Chapter~9.2.1]{Evans}) can be applied to each of
  these equations in order to find their solutions by iteration.

  This theorem is applicable whenever $F_n$ is contractive on a closed subset
  $T$ of $\mathbb C$ and $T$ is invariant under $F_n.$ Due to the fact that the
  derivative of $F_n$ satisfies
  \begin{equation*}
 	 |F_n'|=\frac{1}{a\sqrt{2\lambda}}\frac{1}{|1+\tilde\kappa^2|^{1/2}}\leq\frac{1}{a\sqrt{2\lambda}}\frac{1}{|1-|\tilde\kappa|^2|^{1/2}},
  \end{equation*}
  the map is contractive on $T=\{\tilde\kappa\in\mathbb C\big||\tilde\kappa|\leq K\}$ with $K$ being a
  constant which satisfies
  \begin{equation*}
	0<K<\sqrt{1-\frac{1}{2a^2\lambda}}.
  \end{equation*}
  The requirement of having an invariant subset will restrict $K$ further. In
  order for $|F_n(\tilde\kappa)|\leq K$ to hold, we need
  \begin{equation*}
	|F_n(\tilde\kappa)| \leq |F_n(\tilde\kappa)-F_n(0)|+|F_n(0)|
	\leq\frac{1}{a\sqrt{2\lambda}}\frac{|\tilde\kappa|}{|1-|\tilde\kappa|^2|^{1/2}}+|n|\frac{\pi}{2a\sqrt{2\lambda}}\leq K.
  \end{equation*}
  But since every term is positive, this can only be the case for non-zero
  $\tilde\kappa,$ if
  \begin{equation*}
	|n|\frac{\pi}{2a\sqrt{2\lambda}}<K.
  \end{equation*}
  Together with the upper bound on $K$ from above, we thereby conclude that
  \begin{equation}\label{eq:lem_proof}
	1+\left(\frac{n\pi}{2}\right)^2<2a^2\lambda
  \end{equation}
  needs to hold true for such a $K$ to exist. Thus, for fixed parameters $a$ and
  $\lambda$ the Banach Fixed Point Theorem can only be applied to those
  fixed-point equations whose index satisfies~\eqref{eq:lem_proof}. If the
  parameters are such that this inequality is fulfilled, $n=0$ will always be allowed.
  But the corresponding fixed-point equation has $\tilde\kappa=0$ as its unique
  solution, which leads us to the trivial solution of the Schrödinger equation.
  Since this was excluded in the definition of Gamow functions, we require
  $|n|\geq1.$

  To prove the second part of the Lemma let $a>0$ be fixed, assume that
  $\lambda$ is big enough for~\eqref{eq:Gamow_Condition} to allow resonances and
  fix an $n$ which satisfies this condition. Then we can iterate the
  corresponding fixed-point equation in order to find the location of the $n$th 
  resonance. If this is done up to second order starting with
  $\tilde\kappa^{(0)}=0,$ we end up with a rather cumbersome expression.
  Expanding this expression in powers of $\lambda^{-1/2}$ and keeping only the
  first non-vanishing order of real and imaginary part yields
  \begin{equation*}
	\tilde\kappa^{(2)}_n=\frac{n\pi}{2a\sqrt{2\lambda}}-i\frac{n\pi}{4a^2\lambda}+O(\lambda^{-3/2})\quad\text{and}\quad
	\tilde z_n=\frac{n\pi}{2a}-i\frac{n\pi}{2a^2\sqrt{2\lambda}}+O(\lambda^{-1})
  \end{equation*}
  respectively. If we use the definition of $\tilde z$ and expand again we also
  get
  \begin{equation*}
	z_n=(\tilde z_n^2+2\lambda)^{1/2}=
	\left(2\lambda+\frac{n^2\pi^2}{4a^2}\right)^{1/2}-i\frac{n^2\pi^2}{4a^3\sqrt{2\lambda}}\left(2\lambda+\frac{n^2\pi^2}{4a^2}\right)^{-1/2}+O(\lambda^{-3/2}),
  \end{equation*}
  where we implicitly used that $\Im \sqrt z<0$ whenever $\Im z<0.$ And in order
  for $\Im\tilde z_n$ to be negative, $n$ is restricted to positive integers. 
\end{proof}

To determine the Gamow functions that correspond to the resonances just found in
the previous lemma, it remains to solve the homogeneous system of linear
equations~\eqref{eq:Gamow_LGS}. A series of straightforward calculations yields
\begin{align*}
  G(x) = B
	\begin{cases}
	  \frac{2\tilde z}{\tilde z - z} e^{-i\tilde za}e^{-iz(x+a)} &,x<-a\\\\
	  e^{i\tilde zx}+\frac{\tilde z - z}{\tilde z + z}e^{i\tilde za}e^{-i\tilde z(x-a)} &,|x|\leq a\\\\
	  \frac{2\tilde z}{\tilde z + z} e^{i\tilde za}e^{iz(x-a)} &,x>a.
	\end{cases}
\end{align*}
Evaluating this expression at $z=z_n$ immediately implies that
\begin{align}
  G_n(x) = 
  \begin{cases}
	  \frac{\tilde z_n}{\tilde z_n + z_n} e^{i\tilde z_na}e^{-iz_n(x+a)} &,x<-a\\\\
	  \cos \tilde z_nx &,|x|\leq a\\\\
	  \frac{\tilde z_n}{\tilde z_n + z_n} e^{i\tilde z_na}e^{iz_n(x-a)} &,x>a
	\end{cases}
\end{align}
for odd $n,$ while
\begin{align}
  G_n(x) = 
  \begin{cases}
	  -\frac{\tilde z_n}{\tilde z_n + z_n} e^{i\tilde z_na}e^{-iz_n(x+a)} &,x<-a\\\\
	  i\sin \tilde z_nx &,|x|\leq a\\\\
	  \frac{\tilde z_n}{\tilde z_n + z_n} e^{i\tilde z_na}e^{iz_n(x-a)} &,x>a
	\end{cases}
\end{align}
for even $n,$ where $B$ was set to $1/2.$ This can be seen by using
\begin{align*}
  e^{i2\tilde z_na}\frac{\tilde z_n - z_n}{\tilde z_n + z_n}=
  \begin{cases}
	+1\quad,n\text{ odd}\\
	-1\quad,n\text{ even},
  \end{cases}
\end{align*}
which directly follows from equation~\eqref{eq:Gamow_Condition_temp} with
$-in2\pi$ added to the exponent on the left hand side.

\begin{figure}[!ht]
  \centering
  \includegraphics[scale=0.6]{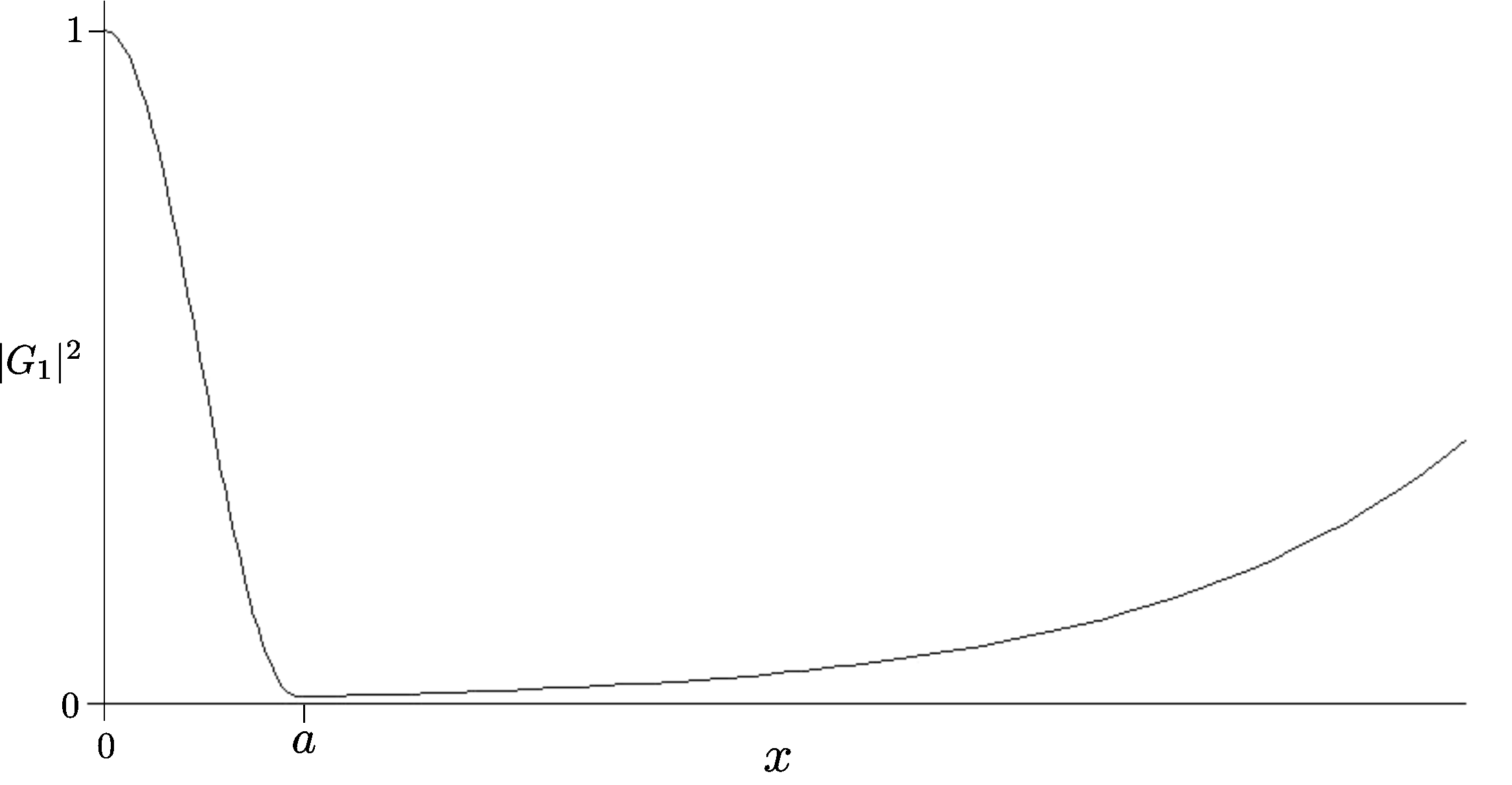}
  \caption{Plot of the squared modulus $|G_1|^2$ of the Gamow function
  corresponding to the first resonance with parameter choice $a=2$ and
  $\lambda=30.$ Notice that for $x>a$ the function is strongly compressed in the
  $x$-direction, to make the exponential increase better visible. This was done
  by inserting the factor $\lambda$ in front of $(x-a).$}
  \label{fig:Gamow}
\end{figure}

We conclude this chapter with Figure~\ref{fig:Gamow}, which illustrates that the
Gamow functions do indeed increase exponentially in $x.$ In the next chapter
we will establish a connection between these functions and the generalized
eigenfunctions. For this purpose, the explicit expression for the resonances will
become relevant again, since we will discover that the generalized
eigenfunctions have poles located at the resonances.


%% file: module32.tex





\chapter{Generalized Eigenfunctions and Strategy}
\label{ch:Eigenfunctions}

In this chapter we will establish precisely how the Gamow functions are connected
with the generalized eigenfunctions of the potential $V_\lambda.$ Thereby, the
relation
\begin{equation*}
  \phi(k,x)\sim\eta(k)G(x)
\end{equation*}
that was announced in the introduction will be turned into a rigorous statement,
from which the function $\eta$ will be deduced. But the major part of this
chapter is devoted to the properties of $\eta$ that are needed to prove the
exponential decay of truncated Gamow functions within a certain time regime.

\section{Calculation of the Eigenfunctions}

First of all, the generalized eigenfunctions that belong to the potential
$V_\lambda$ will be calculated. Clearly, there are symmetric and anti-symmetric
solutions to a symmetrical potential and considering both cases separately
simplifies calculations considerably. An ansatz for the symmetric generalized
eigenfunctions is given by
\begin{align*}
  \phi(k,x) =
  \begin{cases}
	Ae^{ikx}+Be^{-ikx}\quad &, x<-a\\
	C(e^{i\tilde kx}+e^{-i\tilde kx}) &, |x| \leq a\\
	Be^{ikx}+Ae^{-ikx} &, x>a,
  \end{cases}
\end{align*}
where $\tilde k=\sqrt{k^2-2\lambda}$. Due to the symmetry condition, it is
sufficient to require that $\phi$ and its first derivative are continuous at $x$
equal to $+a$ in order to guarantee continuity everywhere. And this yields the
following homogeneous system of linear equations
\begin{align*}
	Ae^{-ika}+Be^{ika}-C(e^{i\tilde ka}+e^{-i\tilde ka}) &= 0\\
	-Ae^{-ika}+Be^{ika}-\frac{\tilde k}{k}C(e^{i\tilde ka}-e^{-i\tilde ka}) &=0.
\end{align*}
Upon solving this system, the symmetric generalized eigenfunctions are given by
\begin{align}\label{eq:GEF_Symmetric}
  &\phi(k,x)=Ae^{-ika}
  \begin{cases}
	e^{ik(x+a)}+\frac{\bar F(k)}{F(k)}e^{-ik(x+a)}\quad &,x<-a\\\\
	\frac{2}{F(k)}\cos\tilde kx &,|x|\leq a\\\\
	e^{-ik(x-a)}+\frac{\bar F(k)}{F(k)}e^{ik(x-a)} &,x>a
  \end{cases}\\
  &\text{with}\quad F(k)=\cos\tilde ka-i\frac{\tilde k}{k}\sin\tilde ka.\nonumber
\end{align}
Since this is a $L^\infty$-solution to the Schrödinger equation, it can not be
normalized in the usual~$L^2$-sense. But similarly to the usual Fourier
transform it is required\linebreak[4] that $\int\overline{\phi(k',x)}\phi(k,x)\,dx=\delta(k'-k).$ And this
gives~$A=\frac{1}{\sqrt{4\pi}}$.

Analogously the ansatz for the anti-symmetric generalized eigenfunctions leads
to the following expression
\begin{align}\label{eq:GEF_Antisymmetric}
  &\phi(k,x)=\frac{e^{-ika}}{\sqrt{4\pi}}
  \begin{cases}
	e^{ik(x+a)}-\frac{\bar J(k)}{J(k)}e^{-ik(x+a)}\quad &,x<-a\\\\
	\frac{2i}{J(k)}\sin\tilde kx &,|x|\leq a\\\\
	-e^{-ik(x-a)}+\frac{\bar J(k)}{J(k)}e^{ik(x-a)} &,x>a
  \end{cases}\\
  &\text{with}\quad J(k)=\frac{\tilde k}{k}\cos\tilde ka-i\sin\tilde ka.\nonumber
\end{align}

\section{Refined Strategy}
\label{sec:Strategy}

The purpose of this section is to refine the argument that served to illustrate
in the introduction how the exponential decay of $e^{-iHt}\psi_0$ arises within
a certain time regime.

A vital part of this argument was the relation $\phi(k,x)\sim\eta(k)G(x).$ But
given that the generalized eigenfunctions solve $(\epsilon-H_0)\phi=V\phi$ their
generic structure is
\begin{equation}\label{eq:GEF_Structure}
  \phi(k,x)=e^{ikx}+\Phi(k,x),
\end{equation}
that is general solution to the homogeneous part plus particular solution to the
inhomogeneous part. And usually it is not the $\Phi(k,x)$ part that governs the
time evolution
\begin{equation}\label{eq:Time_Evolution}
  e^{-iHt}\psi_0(x)=\int\mathcal F\psi_0(k)\phi(k,x)e^{-i\frac{k^2}{2}t}\,dk,
\end{equation}
but the $e^{ikx}$ part. However, only $\Phi$ can cause the generalized
eigenfunctions to mimic the shape of the Gamow function $G.$ Therefore, we are
interested in the situation in which this function gives the dominating
contribution to the generalized eigenfunctions. And heuristically this will be
the case when
\begin{equation*}
  |\Phi(k,x)|=|\phi(k,x)-e^{ikx}|\gg|e^{ikx}|.
\end{equation*}

\begin{figure}[t]
  \centering
  \includegraphics[scale=0.6]{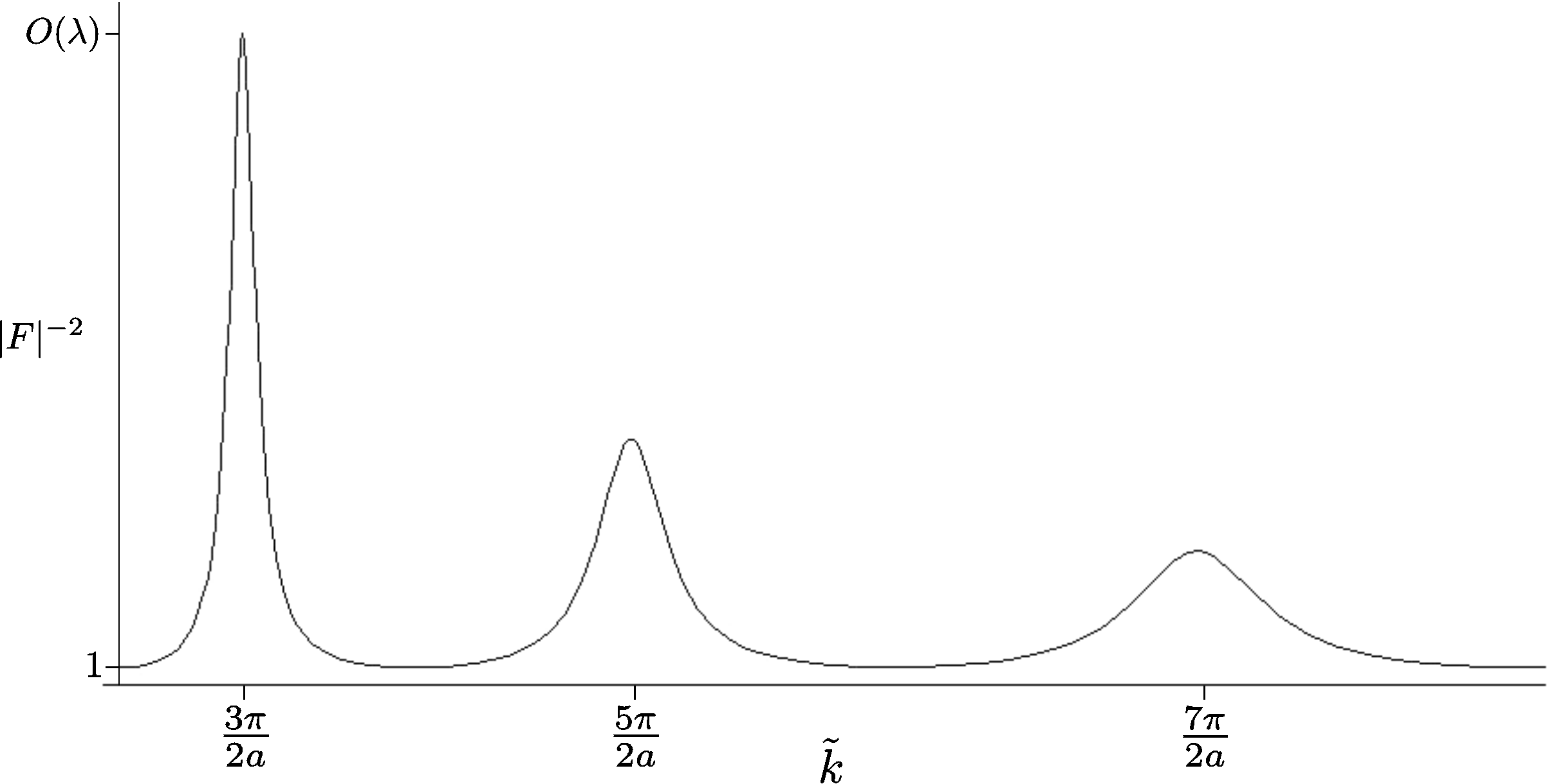}
  \caption{Plot of $|F|^{-2}$ with parameter choice $a=2$ and $\lambda=100,$
  where~$kF(k)$ denotes the Jost function of the symmetric generalized
  eigenfunctions. The $k$-axis was rescaled by using the relation $\tilde
  k=\sqrt{k^2-2\lambda}.$}
  \label{fig:Jost}
\end{figure}

Clearly, the generalized eigenfunctions that belong to $V_\lambda$ have the
structure~\eqref{eq:GEF_Structure}. And according to the above-mentioned
argument, we expect that they take the shape of $G$ when
\begin{equation*}
  |F|^{-1}\gg1\quad\text{and}\quad|J|^{-1}\gg1
\end{equation*}
respectively. In fact, the squared modulus of $F^{-1}$ consists of a series of
Breit-Wigner distributions whose height is of $O(\lambda).$ This is illustrated
in Figure~\ref{fig:Jost} and will be proven rigorously later. Since these
distributions are represented by the formula
\begin{equation*}
  \frac{|a_{-1}|^2}{(k-z'_n)^2+z''^2_n}=\frac{|a_{-1}|^2}{(k-z'_n-iz''_n)(k-z'_n+iz''_n)},
\end{equation*}
their appearance in $|F|^{-2}$ reflects the fact that the continuation of
$\phi(k,x)$ to the complex $k$-plane has poles due to $F^{-1}.$ Therefore, we
just need to apply the calculus of residues to the above-mentioned
integral~\eqref{eq:Time_Evolution} to extract a term being proportional to
$\exp(-z'_nz''_n\,t).$ So for those~$t$ for which this term gives the dominating
contribution to $e^{-iH_\lambda t}\psi_0,$ the wave function will decay exponentially.
But before we pursue this idea further, the existence of the poles~will be
proven in the next lemma.

\begin{lemm}\label{lem:Poles}
  The complex continuations of $F$ and $J$ have roots, which coincide with the
  resonances~$z_n$ corresponding to odd $n$ and even $n$ respectively. In both
  cases the multiplicity of the roots is one.
\end{lemm}
\begin{proof}
  As already noted in Section~\ref{sec:Gamow}, the resonances $z_n$
  satisfy the following identity
  \begin{align*}
    e^{i2a\tilde z_n}\,\frac{\tilde z_n-z_n}{\tilde z_n+z_n}=
    \begin{cases}
	  +1\quad,n\text{ odd}\\
	  -1\quad,n\text{ even}.
	\end{cases}
  \end{align*}
  Using this, a straightforward calculation shows that
  \begin{equation*}
	\frac{e^{ia\tilde z_n}-e^{-ia\tilde z_n}}{e^{ia\tilde z_n}+e^{-ia\tilde	z_n}}=\frac{z_n}{\tilde z_n}
  \end{equation*}
  holds for odd $n.$ And it is not difficult to see that this equation is
  equivalent to $\tan\tilde z_na=-i\frac{z_n}{\tilde z_n},$ which is the equation
  satisfied by the roots of $F.$ An analogous argumentation holds for $J$ with
  even $n.$

  To show that the multiplicity of the roots is one, we need to show that
  $F'(z_{2n-1})$ and $J'(z_{2n})$ do not vanish. In this regard, notice that
  \begin{equation*}
	F'(z_{2n-1})=-ia\left(\cos\tilde ka-i\frac{k}{\tilde k}\sin\tilde ka\right)-i\frac{2\lambda}{\tilde kk^2}\sin\tilde ka\bigg|_{k=z_{2n-1}}.
  \end{equation*}
  Using the fact that $F(z_{2n-1})=0,$ this can be expressed as
  \begin{equation*}
	F'(z_{2n-1})=-\left(a-\frac{i}{k}\right)\frac{2\lambda}{\tilde kk}\sin\tilde ka\bigg|_{k=z_{2n-1}}.
  \end{equation*}
  The term in brackets can not vanish, because $\Re z_{2n-1}$ is non-zero. Since
  the sine has roots only on the real axis and $\Im z_{2n-1}>0,$ it can not
  vanish either. Hence, $F'(z_{2n-1})$ is non-zero. An analogous
  argumentation holds for $J.$
\end{proof}

According to the previous lemma, the poles of the generalized eigenfunctions
coincide with the resonances, which justifies that the Gamow 'eigenvalues' have
been termed this way. In order to see this notice that $kF(k)$ and $kJ(k)$ are so
called Jost functions (see~\cite[Section~2]{Skibsted86}), which define the
Scattering-Matrix by
\begin{equation*}
  S(k)=\frac{\bar F(k)}{F(k)}\quad\text{and}\quad S(k)=\frac{\bar J(k)}{J(k)}
\end{equation*}
respectively. Since it is common to define resonances as poles of the complex
continuation of this operator, we see that they coincide with the roots of $F$ and
$J$ and thereby with the Gamow 'eigenvalues'. Although, this was only verified
for $V_\lambda,$ Skibsted's article~\cite{Skibsted86} shows that it remains true
for more general potentials. 

Another consequence of this observation, is that the Gamow function $G_n$ and
the residue of~$\phi(k,\cdot)$ at $z_n$ are linearly dependent. In order to see
this let $\varphi$ be defined by
\begin{equation*}
  \phi(k,\cdot)=\frac{\varphi(k,\cdot)}{H(k)},
\end{equation*}
where $H$ denotes either $F$ or $J$. Then
$\mathrm{Res}\,\,\phi=a_{-1}\,\varphi(z_n,\cdot).$ And given that
the Jost function evaluated at $z_n$ can be expressed as
(see~\cite[Section~2]{Skibsted86})
\begin{equation*}
  z_nH(z_n)=
  \begin{vmatrix}
  	G_n(x) 				& \varphi(z_n,x)\\
	\frac{d}{dx}G_n(x) 	& \frac{d}{dx}\varphi(z_n,x)
  \end{vmatrix},
\end{equation*}
the linear dependence becomes immediately evident, when we recall that
$H(z_n)=0.$ Therefore, we can establish a precise relation between the
generalized eigenfunctions and the Gamow functions by expanding $\phi$ in a
Laurent series about the resonance $z_n.$ This yields the expression
\begin{equation*}
  \phi(k,x)=\underbrace{\frac{a_{-1}}{k-z_n}\,c}_{\eta(k)}\,G_n(x)+\text{Remainder},
\end{equation*}
from which $\eta(k)$ can be simply read off.

\section{Laurent Expansion of the Eigenfunctions}

This section will show that for big enough $\lambda$ there is an interval
$[z'_n-\delta,z'_n+\delta]$ on which the principal part $\eta(k)G$ of the
Laurent expansion of $\phi(k,\cdot)$ is a good approximation for the generalized
eigenfunctions. In the next chapter this will allow us to justify the
approximation
\begin{equation*}
  e^{-iH_\lambda t}\psi_0=\int\mathcal F\psi_0(k)\phi(k,\cdot)e^{-i\frac{k^2}{2}t}\,dk
  \sim G(\cdot)\,\int\mathcal F\psi_0(k)\eta(k)e^{-i\frac{k^2}{2}t}\,dk
\end{equation*}
for those initial wave functions, whose energy distribution $\mathcal F\psi_0$ is
concentrated in the above-mentioned interval.

Due to the fact that $V_\lambda$ has compact support, the generalized eigenfunctions
break up in a part being valid on the support of $V_\lambda$ and a part being
valid on its complement. Therefore, it is convenient to consider the time
evolution integral~\eqref{eq:Time_Evolution} on these subsets of $\mathbb R$
separately. For now, we focus on $supp\,\,V_\lambda.$ And this will require
frequent use of the characteristic function~$\chi_{_{[-a,a]}},$ which is why it
is advantageous to introduce the following shorthand
\begin{equation*}
  \chi_a:=\chi_{_{[-a,a]}}.
\end{equation*}
Moreover, it is convenient to restrict ourselves to symmetric initial wave
functions such that only the symmetric generalized
eigenfunctions~\eqref{eq:GEF_Symmetric} contribute to the
integral~\eqref{eq:Time_Evolution}. However, every result has a straightforward
extension to the anti-symmetric case.

\begin{lemm}\label{lem:Laurent}
  Let $\phi$ denote the symmetric generalized eigenfunctions of $H_\lambda$ with
  $\lambda\in\mathbb R^+$ and let~$z_{2n-1}$ with~$n\geq 1$ denote its
  resonances. Then $\phi$ can be expanded in the Laurent series
  \begin{equation}\label{eq:GEF_Laurent}
	\phi(k,x)=e^{-ika}\left(\frac{a_{-1}(x)}{k-z_{2n-1}}+a_0(x)+R_0(k,x)\right),
  \end{equation}
  for every $x$ in $[-a,a].$ And if $\lambda$ is chosen big enough, there is a
  $\delta>0,$ such that this expansion converges absolutely and uniformly on
  $[z'_{2n-1}-\delta,z'_{2n-1}+\delta],$ provided $n$ satisfies
  \begin{equation*}
	1\leq n<\sqrt{\frac{\lambda}{2}-\sqrt{2\lambda}}=:n_\lambda.
  \end{equation*}
\end{lemm}

\begin{proof}
  In Lemma~\ref{lem:Poles} it was proven that the resonances $z_{2n-1}$ are first
  order poles of the complex continuation of $\phi(k,x)$ to the lower
  complex $k$ plane. Thus, it is clear that the Laurent expansion of $\phi$ is
  given by~\eqref{eq:GEF_Laurent}. In order to prove that it converges in the
  above-mentioned interval about $z'_{2n-1},$ we will determine its annulus of
  convergence. Since the resonances are separated points in the complex plane,
  it is clear that it will be a punctured disc around that particular resonance
  $z_{2n-1}.$ So it remains to estimate the outer radius of convergence, which can be
  done by estimating the distance between neighboring poles of $\phi.$

  By Lemma~\ref{lem:Poles}, the symmetric generalized eigenfunctions have poles
  numbered by odd indices, while the anti-symmetric ones have poles numbered by
  even indices. Therefore, we need to ensure that
  \begin{equation}\label{eq:proof_theorem1}
	|z_{2n+1}-z_{2n-1}|^2>4z''^2_{2n+1}>4z''^2_{2n-1}\qquad\text{for }n\geq1,
  \end{equation}
  where the last inequality will automatically hold if the first one is true.
  This is due to $z''^2_k$ being a monotonically increasing function of $k.$
  Notice, that the factor in front of the imaginary part is chosen for
  convenience, but is not optimal. Using the explicit formula for the resonances
  given in Lemma~\ref{lem:Resonances}, we find
  \begin{align*}
	|z_{2n+1}-z_{2n-1}|^2&=\left(\sqrt{2\lambda+\frac{(2n+1)^2\pi^2}{4a^2}}-\sqrt{2\lambda+\frac{(2n-1)^2\pi^2}{4a^2}}\right)^2\\
	&\qquad+\frac{\pi^2}{4a^2}\frac{1}{2\lambda}\left(\frac{2n+1}{\left(1+\frac{4a^2\,2\lambda}{(2n+1)^2\pi^2}\right)^{1/2}}
	-\frac{2n-1}{\left(1+\frac{4a^2\,2\lambda}{(2n-1)^2\pi^2}\right)^{1/2}}\right)^2.
  \end{align*}
  By exploiting the formula $(a+b)(a-b)=a^2-b^2,$ the term in brackets appearing
  in the first summand can be simplified to
  \begin{align*}
	\frac{\pi^2}{4a^2}\frac{8n}{\sqrt{2\lambda+\frac{(2n+1)^2\pi^2}{4a^2}}+\sqrt{2\lambda+\frac{(2n-1)^2\pi^2}{4a^2}}}
	\geq\frac{\pi^2}{4a^2}\frac{8n}{2\sqrt{2\lambda+\frac{(2n+1)^2\pi^2}{4a^2}}}.
  \end{align*}
  The second summand can be simplified by using a similar argumentation, which
  yields the following lower bound
  \begin{align*}
	\frac{\pi^2}{4a^2}\frac{1}{2\lambda}\left(\frac{2}{\left(1+\frac{4a^2\,2\lambda}{(2n+1)^2\pi^2}\right)^{1/2}}\right)^2
	=\left(\frac{\pi^2}{4a^2}\right)^2\frac{1}{2\lambda}\,\frac{4(2n+1)^2}{2\lambda+\frac{(2n+1)^2\pi^2}{4a^2}}.
  \end{align*}
  Together these lower bounds imply that inequality~\eqref{eq:proof_theorem1} is
  full filled, if \begin{align*}
	\left(\frac{\pi^2}{4a^2}\right)^2\frac{(4n)^2+\frac{1}{2\lambda}4(2n+1)^2}{2\lambda+\frac{(2n+1)^2\pi^2}{4a^2}}>4z''^2_{2n+1}
	=\left(\frac{\pi^2}{4a^2}\right)^2\frac{1}{2\lambda}\frac{4(2n+1)^4}{2\lambda+\frac{(2n+1)^2\pi^2}{4a^2}}.
  \end{align*}
  Hence, inequality~\eqref{eq:proof_theorem1} holds if
  \begin{equation*}
	(4n)^2>\frac{2}{\lambda}(2n+1)^2((2n+1)^2-1).
  \end{equation*}
  Some simple rearrangements and estimates yield the following inequality
  \begin{equation*}
	0>n^2-\left(\frac{\sqrt{2\lambda}}{2}-1\right)n+\frac{1}{4},
  \end{equation*}
  where equality would hold for
  \begin{equation*}
	n_\pm=\frac{1}{2}\left(\left(\frac{\sqrt{2\lambda}}{2}-1\right)\pm\sqrt{\left(\frac{\sqrt{2\lambda}}{2}-1\right)^2-1}\right).
  \end{equation*}
  Clearly, $n_-$ will approach zero in the limit of $\lambda$ tending to
  $\infty.$ Therefore, only $n_+$ is of relevance when $\lambda\gg1$ and
  $n\geq1.$ And if we regard $n_+$ as arithmetic mean, this allows us to conclude
  that the distance $|z_{2n+1}-z_{2n-1}|$ between neighboring resonances is
  larger than $2|z''_{2n+1}|$ for all integer~$n,$ which satisfy
  \begin{equation*}
	1\leq n<\sqrt{\frac{\lambda}{2}-\sqrt{2\lambda}}=:n_\lambda.
  \end{equation*}
  So for these resonances there is a non-zero positive $\delta,$ such that the
  Laurent expansion~\eqref{eq:GEF_Laurent} converges absolutely and uniformly
  for all $k$ in $[z'_n-\delta,z'_n+\delta].$ 
\end{proof}

That the region in which the Laurent expansion~\eqref{eq:GEF_Laurent} converges
ceases to intersect the real axis is not surprising. It reflects the fact that
the influence of the resonances fades the larger their real part grows. From a
physical point of view this can be anticipated, since initial wave functions
with high energy $\left<\psi_0,H_\lambda\psi_0\right>$ impinge on the 'barrier'
with increased rate. Thus, their lifetime on top of the plateau is reduced.

But more importantly, Lemma~\ref{lem:Laurent} shows that
\begin{equation*}
  \eta_n(k)=e^{-ika}\frac{a_{-1}}{k-z_{2n-1}}\quad\text{with}\quad
  a_{-1}=\res{z_{2n-1}}{\frac{1}{\sqrt\pi F(k)}},
\end{equation*}
which verifies the Breit-Wigner shape of $|\eta_n|^2$ as announced in the
introduction.

The estimates on the remainder of the Laurent expansion, will involve contour
integrals around circles in the complex $k$-plane. But to handle them some
properties of the corresponding contour performed by
$\tilde k=\sqrt{k^2-2\lambda}$ are needed. And the relevant details will be
established in the following lemma by considering the function
$\sqrt{c+d\,e^{i\varphi}},$ where $0\leq d\leq c$ are real constants. Basically,
this function will perform a contour which looks like an egg whose symmetry axis
coincides with the real axis. This becomes plausible by rewriting the argument
of the complex square root in polar form $r(\varphi)\,\exp(iS(\varphi)).$ Then the
effect of the square root on the exponential is to compress the circle in the
direction of the imaginary axis, whereas its effect on the radius $r(\varphi)$
shifts its center and will cause the circle to be compressed along the real
axis. Since the strength of these effects depends on $\varphi,$ the resulting
contour will roughly look like an egg. For the purpose of estimating the
remainder, the only information we need is in which region of the complex
$k$-plane the contour performed by $\tilde k$ can be found. And the preceding
illustration suggests that it is an annulus. The following lemma makes this
precise.

\begin{lemm}\label{lem:Egg}
	Let $0\leq d\leq c$ be real and let $\varphi$ lie in $[0,2\pi).$ Then the
	complex contour $\sqrt{c+d\,e^{i\varphi}},$ remains within the
	annulus $S$ enclosed by the circles
	\begin{align*}
	  z_{out}&=\sqrt c+r_{out}\,e^{i\varphi} \text{ with } r_{out}=\sqrt c-\sqrt{c-d}\text{ and}\\
	  z_{in}&=\sqrt c+r_{in}\,e^{i\varphi} \text{ with } r_{in}=\sqrt{c+d}-\sqrt c.
	\end{align*}
\end{lemm}

\begin{proof}
  The assertion follows, if we show that the continuous function $f(\varphi)$
  defined\linebreak[4] by $|\sqrt{c+d\,e^{i\varphi}}-\sqrt c|$ remains within the interval
  $[r_{in},r_{out}].$ This can be done by showing that~$f^2$ takes its only
  maximum at $\varphi$ being equal to $\pi$ and its only minimum at $\varphi$
  being equal to zero. In order to do that, it is enough to show that $f^2$ is
  monotonically increasing on the interval $(0,\pi),$ while it is monotonically
  decreasing on $(\pi,2\pi).$

  The monotonicity can be deduced from the sign of the first derivative of
  $f^2.$ But this will be easier if $f^2$ is rewritten first. By using the
  equality $(a+b)(a-b)=a^2-b^2,$ we see that
  \begin{equation*}
	f^2(\varphi)=\frac{d^2}{|\sqrt{c+d\,e^{i\varphi}}+\sqrt c|^2},
  \end{equation*}
  which has the following derivative with respect to $\varphi$
  \begin{equation*}
	-\frac{d^2}{|\sqrt{c+d\,e^{i\varphi}}+\sqrt c|^4}
	\partial_\varphi|\underbrace{\sqrt{c+d\,e^{i\varphi}}+\sqrt c|^2}_{=:\tilde f^2}.
  \end{equation*}
  That means, $f^2$ is increasing whenever $\tilde f^2$ is decreasing and vice
  versa. In order to determine which of these cases occurs for which $\varphi,$
  we need to carry out the differentiation of $\tilde f^2$ symbolically
  \begin{align*}
	\partial_\varphi\tilde f^2&=2\,(\Re\sqrt{c+d\,e^{i\varphi}}+\sqrt c)\,\partial_\varphi\Re\sqrt{c+d\,e^{i\varphi}}\\
	&\qquad+2\,\Im\sqrt{c+d\,e^{i\varphi}}\,\partial_\varphi\Im\sqrt{c+d\,e^{i\varphi}}.
  \end{align*}
  With the help of the formula
  $\sqrt{x+iy}=2^{-1/2}\left((\sqrt{x^2+y^2}+x)^{1/2}+i\,\text{sgn}(y)(\sqrt{x^2+y^2}-x)^{1/2}\right),$
  which is easily proved by rewriting the argument of the square root in polar
  form, we can bring real and imaginary part in the following form
  \begin{align*}
 	\Re\sqrt{c+d\,e^{i\varphi}}&=\frac{1}{\sqrt 2}\left(\sqrt{c^2+d^2+2cd\cos\varphi}+c+d\cos\varphi\right)^{1/2}\text{ and}\\
	\Im\sqrt{c+d\,e^{i\varphi}}&=\text{sgn}(\sin\varphi)\frac{1}{\sqrt 2}\left(\sqrt{c^2+d^2+2cd\cos\varphi}-(c+d\cos\varphi)\right)^{1/2}.
  \end{align*}
  Using these formulas, the real and imaginary part can be differentiated with
  respect to $\varphi,$ yielding
  \begin{align*}
	\partial_\varphi\Re\sqrt{c+d\,e^{i\varphi}}&=\frac{1}{4}\frac{1}{\Re\sqrt{c+d\,e^{i\varphi}}}\left(\frac{-cd\sin\varphi}{\sqrt{c^2+d^2+2cd\cos\varphi}}-d\sin\varphi\right)\text{ and}\\
	\partial_\varphi\Im\sqrt{c+d\,e^{i\varphi}}&=\frac{1}{4}\frac{\text{sgn}(\sin\varphi)}{\Im\sqrt{c+d\,e^{i\varphi}}}\left(\frac{-cd\sin\varphi}{\sqrt{c^2+d^2+2cd\cos\varphi}}+d\sin\varphi\right).
  \end{align*}
  Therefore, we arrive at the following expression for the derivative of $\tilde f^2$
  \begin{align*}
	\partial_\varphi\tilde f^2&=
	-\frac{d\sin\varphi}{2}\left(\frac{c}{\sqrt{c^2+d^2+2cd\cos\varphi}}+1\right)\left(1+\frac{\sqrt c}{\Re\sqrt{c+d\,e^{i\varphi}}}\right)\\
	&\quad+\frac{d\sin\varphi}{2}\left(\frac{-c}{\sqrt{c^2+d^2+2cd\cos\varphi}}+1\right)\text{sgn}(\sin\varphi)\\
	&=g(\varphi)+h(\varphi).
  \end{align*}
  Comparing both summands factor by factor, we see that $|g|$ is strictly bigger
  than $|h|$ on the interval $(0,\pi)$ and $(\pi,2\pi).$ And the signs of $g$
  and $h$ are such that $f^2$ is monotonically increasing in the first case,
  while it is monotonically decreasing on $(\pi,2\pi).$ Therefore, $f^2$ reaches
  its maximum at $\pi$ and its minimum at zero. So if we define
  \begin{align*}
	r_{out}&=f(\pi)=\sqrt c-\sqrt{c-d}\text{ and}\\
	r_{in}&=f(0)=\sqrt{c+d}-\sqrt c,
  \end{align*}
  we can conclude that the contour performed by $\sqrt{c+d\,e^{i\varphi}}$ remains within the
  annulus $S,$ which is enclosed by the circles with radii $r_{in}$ and
  $r_{out}$ both being centered at $\sqrt c.$
\end{proof}

We can now estimate the modulus of the remainder of the Laurent expansion.

\begin{thm}\label{thm:Laurent_Coefficients}
  Let $\phi$ denote the symmetric generalized eigenfunctions of $H_\lambda$ and
  let $n$ be a fixed odd integer satisfying $1\leq n < n_\lambda.$ If $\lambda$
  is chosen big enough, the Laurent expansion of~$e^{ika}\phi(k,\cdot)$~in~$k$
  about the resonance $z_n$ converges for all $k$ in $[z'_n-\delta,z'_n+\delta]$
  with
  \begin{equation*}
  	\delta = \frac{1}{2}(z'_n-z'_{n-1}).
  \end{equation*}
  Furthermore, the Laurent coefficients $a_m(\cdot)$ and the remainder $R_0(k,\cdot)$ satisfy
  \begin{align*}
	&\|\chi_aa_m(\cdot)\|_\infty\leq O(\lambda^{m/2})\text{ and}\\
	&\|\chi_aR_0(k,\cdot)\|_\infty\leq O(\lambda^{1/2})|k-z_n|.
  \end{align*}
\end{thm}

\begin{proof}
  The $L^\infty$ bounds on the coefficients will be proven first.
  Since~\eqref{eq:GEF_Laurent} is a Laurent expansion its coefficients can be
  calculated by
  \begin{equation*}
	a_m(\cdot)=\frac{1}{2\pi i}\oint_C\frac{\phi(z,\cdot)e^{iza}}{(z-z_n)^{m+1}}\,dz,
  \end{equation*}
  where $C$ is a closed contour around the pole $z_n,$ which remains within the
  annulus of convergence. Using the explicit form of the generalized
  eigenfunctions given in~\eqref{eq:GEF_Symmetric}, we therefore have the
  following upper bound
  \begin{equation}\label{eq:proof_thm}
	\|\chi_aa_m(\cdot)\|_\infty\leq\frac{1}{\sqrt\pi}\sup_{z\in C}\frac{1}{|\cos a\tilde z-i\frac{\tilde z}{z}\sin a\tilde z|}\,\sup_{z\in C}\,\sup_{x\in\mathbb R}|\chi_a\cos \tilde zx|\,\,\,r^{-m}.
  \end{equation}
  Here $C$ was chosen to be the circle
  \begin{align}\label{circle}
	&\{z_n+re^{i\varphi}\,|\,\varphi\in[0,2\pi)\}\text{ with}\\
	&r=z'_n-z'_{n-1}\nonumber,
  \end{align}
  which is a contour within the annulus of convergence for all $n,$ since only
  one pole is enclosed. To handle the suprema appearing in~\eqref{eq:proof_thm},
  we need to know how the circle $C$ in the $z$ plane looks like in the $\tilde
  z$ plane. Due to the fact that we are only interested in the limit of
  $\lambda$ tending to $\infty,$ we can exploit the dependence of $z_n$ on
  $\lambda$ in order to find
  \begin{align*}
	\tilde z=(z^2-2\lambda)^{1/2}&=(z_n^2-2\lambda+2z_nre^{i\varphi}+r^2e^{i2\varphi})^{1/2}\\
	&=(\tilde z'^2_n+2(z'_n+iz''_n)re^{i\varphi})^{1/2}\\
	&=\left(\tilde z'^2_n+2z'_n\sqrt{2\lambda}\left(1+\frac{1}{2}\frac{(2n-1)^2\pi^2}{4a^22\lambda}-1-\frac{1}{2}\frac{(2n-2)^2\pi^2}{4a^22\lambda}\right)e^{i\varphi}\right)^{1/2}\\
	&=(\tilde z'^2_n+(\tilde z'^2_n-\tilde z'^2_{n-1})e^{i\varphi})^{1/2}=:\sqrt{\tilde z^{'2}_n+r'e^{i\varphi}},
  \end{align*}
  which does not depend on $\lambda$ anymore. Thus,
  \begin{align}\label{eq:proof_thm_cosh}
	\sup_{z\in C}\sup_{x\in\mathbb R}|\chi_a\cos\tilde zx|\leq\sup_{z\in C}\,\sup_{x\in\mathbb R}\,\chi_a\cosh{\tilde z''x} = \sup_{z\in C}\,\cosh{\tilde z''a}\leq c,
  \end{align}
  since $\tilde z''$ is independent of $\lambda$ on the circle $C.$ And for
  similar reasons
  \begin{align*}
	\frac{|\tilde z|}{|z|}|\sin a\tilde z|\leq\frac{c}{|z|}=O(\lambda^{-1/2}),
  \end{align*}
  from which we can deduce
  \begin{align*}
	\sup_{z\in C}\frac{1}{|\cos a\tilde z-i\frac{\tilde z}{z}\sin a\tilde z|}
	&\leq\left(\inf_{z\in C}|\cos a\tilde z|-|\frac{\tilde z}{z}\sin a\tilde z|\right)^{-1}
	\leq\left(\inf_{z\in C}|\cos a\tilde z|-\frac{1}{4}\right)^{-1},
  \end{align*}
  provided $\lambda$ is big enough. So it remains to estimate the infimum of
  $|\cos a\tilde z|$ on the circle $C.$ Regarding this, the information given in
  Lemma~\ref{lem:Egg} will be sufficient, due to the maximum modulus principle
  of complex analysis (see e.g.~\cite{Lang}). According to this principle, the
  modulus of an analytic non-constant function considered on a connected open
  subset $S$ of $\mathbb C,$ reaches its maximum on the boundary $\partial S.$
  Lemma~\ref{lem:Egg} proves that the contour $C$ remains within the annulus $S$
  bounded by the circles
  \begin{align*}
	\tilde z_{in}&=\tilde z'_n+\left(\sqrt{2\tilde z'^2_n-\tilde z'^2_{n-1}}-\tilde z'_n\right)\,e^{i\varphi}\text{ and}\\
	\tilde z_{out}&=\tilde z'_n+(\tilde z'_n-\tilde z'_{n-1})\,e^{i\varphi}.
  \end{align*}
  Since these circles enclose exactly one root of $\cos a\tilde z$ at their
  center, $(\cos a\tilde z)^{-1}$ is analytic within the annulus $S.$ Thus the
  maximum modulus principle is applicable and it can be concluded
  that~$|\cos a\tilde z|$ reaches its minimum on one of the circles
  $\tilde z_{in}$ or $\tilde z_{out}.$

  In order to find the minimum of $|\cos a\tilde z|$ on the above mentioned
  circles, it is easiest to consider the squared modulus. Since $a\tilde z'_n$
  is a root of the cosine, the squared modulus can be rewritten as follows
  \begin{equation*}
	|\cos(a\tilde z'_n+are^{i\varphi})|^2
	=|\sin(are^{i\varphi})|^2
	=\frac{1}{2}(\cosh(2ar\sin\varphi)-\cos(2ar\cos\varphi)),
  \end{equation*}
  where $r$ denotes either $r_{in}$ or $r_{out}.$ Due to the symmetry of this
  expression about $\pi,$ it is enough to consider
  \begin{equation*}
	\partial_\varphi|\sin(are^{i\varphi})|^2=ar(\sinh(2ar\sin\varphi)\cos\varphi-\sin(2ar\cos\varphi)\sin\varphi)
  \end{equation*}
  on the interval $[0,\pi].$ For $\varphi$ in $(0,\pi/2)$ we also know that
  $\cos\varphi$ and thereby the first summand is positive. In this case
  \begin{equation*}
	\partial_\varphi|\sin(are^{i\varphi})|^2\geq ar\sin\varphi(2ar\cos\varphi-\sin(2ar\cos\varphi))> 0,
  \end{equation*}
  since $\sinh x$ is bigger than $x$ and $(x-\sin x)$ is bigger than zero for all
  positive $x.$ However, should~$\varphi$ lie in $(\pi/2,\pi)$ we know that
  $\cos\varphi$ is negative. Then
  \begin{align*}
	\partial_\varphi|\sin(are^{i\varphi})|^2&=-ar(\sinh(-2ar\sin\varphi)\cos\varphi-\sin(-2ar\cos\varphi)\sin\varphi)\\
	&\leq-ar\sin\varphi((-2ar\cos\varphi)-\sin(-2ar\cos\varphi))<0
  \end{align*}
  for the same reasons as above. The monotonicity implied by these
  considerations shows\linebreak[4] that $|\sin(are^{i\varphi})|^2$ reaches its minima at
  zero and $\pi.$ But due to symmetry the modulus of this function at zero will
  not differ from its modulus at $\pi,$ which is why
  \begin{equation*}
	\inf_{z\in C}|\cos a\tilde z|\geq\inf_{z\in S}|\cos a\tilde z|
	=\min\left\{|\cos a\tilde z_{in}(0)|,|\cos a\tilde z_{out}(0)|\right\}.
  \end{equation*}
  A lower bound for this minimum can be found by determining lower bounds for
  the cosines. Since~$\tilde z_{in}(0)=\sqrt{2\tilde z'^2_n-\tilde z'^2_{n-1}},$
  we see that
  \begin{align*}
	|\cos(a\tilde z_{in}(0))|^2=\frac{1}{2}\left(1+\cos\left(2a\sqrt{2\tilde z'^2_n-\tilde z'^2_{n-1}}\right)\right)
	=\frac{1}{2}\left(1+\cos\left(2n\pi\sqrt{1-\frac{1}{2n^2}}\right)\right),
  \end{align*}
  where the explicit location of the resonances $\tilde z_n$ was used.
  Considering the square root in the argument of the remaining cosine as
  deviation from $2n\pi,$ where the cosine would take its maximum, it becomes
  clear that
  \begin{align*}
	|\cos(a\tilde z_{in}(0))|^2\overset{n=1}{\geq}\frac{1}{2}(1+\cos(\sqrt 2\pi))
	\geq\frac{1}{2}\left(1+\cos\left(\frac{5}{4}\pi\right)\right)=\frac{1}{2}\left(1-\frac{1}{\sqrt	2}\right).
  \end{align*}
  Using the explicit location of the resonances $\tilde z_n$ again, we also find
  \begin{equation*}
	|\cos(a\tilde z_{out}(0))|^2=|\cos(n\pi)|^2=1.
  \end{equation*}
  And this immediately implies that
  \begin{equation*}
	\inf_{z\in C}|\cos a\tilde z|\geq\sqrt{\frac{1}{2}\left(1-\frac{1}{\sqrt 2}\right)}>\frac{1}{3},
  \end{equation*}
  from which we can conclude
  \begin{equation*}
	\sup_{z\in C}\frac{1}{|\cos a\tilde z-i\frac{\tilde z}{z}\sin a\tilde z|}\leq\left(\inf_{z\in C}|\cos a\tilde z|-\frac{1}{4}\right)^{-1}\leq12
  \end{equation*}
  provided $\lambda$ is big enough.
  
  This was the missing estimate for the upper bound of the $L^\infty$-norm of 
  the Laurent coefficients~$\chi_aa_m(\cdot).$ Using this inequality together with
  inequality~\eqref{eq:proof_thm_cosh}, we can deduce from the usual Cauchy
  estimate~\eqref{eq:proof_thm} for the Laurent coefficients that
  \begin{align*}
	\|\chi_a a_m(\cdot)\|_\infty\leq cr^{-m}&=c(z'_n-z'_{n-1})^{-m}
	=O(\lambda^{m/2}).
  \end{align*}


  It remains to prove the bound on the remainder $R_0$ of the Laurent expansion.
  Multiplying~\eqref{eq:GEF_Laurent} by $e^{ika}$ and subtracting the principal
  part of the Laurent expansion on both sides yields
  \begin{equation*}
	e^{ika}\phi(k,x)-\frac{a_{-1}(x)}{k-z_n}=a_0(x)+R_0(k,x).
  \end{equation*}
  Since the right hand side is a Taylor expansion of the left hand side in the
  complex variable~$k,$ the Lagrange form of the remainder can be used, to get an
  upper bound on the modulus of~$R_0.$ (Notice that, the Lagrange form of the
  remainder only gives a bound in dimensions bigger than one, but it is not
  exact anymore.) So if $k$ is within the radius of convergence of the
  Taylor series, there is a
  \begin{align*}
	&\zeta\in\{z_n+t(k-z_n)|t\in(0,1)\}\quad\text{such that}\\
	&\|\chi_aR_0(k,\cdot)\|_\infty\leq\left\|\chi_a\partial_k[e^{ika}\phi(k,\cdot)](\zeta)\right\|_\infty\,|k-z_n|,
  \end{align*}
  where we exploited that the $a_m$ are equal to the Laurent coefficients of
  $e^{ika}\phi(k,\cdot).$ Using Cauchy's formula, the derivative that appears in the
  above bound can be estimated as follows
  \begin{align*}
	\|\chi_a\partial_k[e^{ika}\phi(k,\cdot)](\zeta)\|_\infty
	&=\frac{1}{2\pi}\left\|\chi_a\oint_C\frac{e^{iza}\phi(z,\cdot)}{(z-\zeta)^2}\,dz\right\|_\infty\\
	&\leq\frac{1}{2\pi}\sup_{z\in C}\sup_{x\in\mathbb R}|\chi_ae^{iza}\phi(z,x)|\,\oint_C\frac{1}{|z-\zeta|^2}\,dz,
  \end{align*}
  where $C$ needs to be a closed contour that contains $\zeta.$ The supremum can
  be estimated as above, when the domain of $\zeta$ is contained in the interior
  of the circle~\eqref{circle}, provided $C$ is chosen to be this contour. If we
  postpone proving that $\zeta$ satisfies this condition, we can conclude that
  the supremum of $\|\chi_ae^{iza}\phi(z,\cdot)\|_\infty$ over all $z\in C$ is
  of $O(\lambda^0).$ And this together with the
  particular form of the contour $C,$ implies
  \begin{align*}
	\|\chi_a\partial_k[e^{ika}\phi(k,\cdot)](\zeta)\|_\infty&\leq\frac{O(\lambda^0)}{2\pi}r\int_0^{2\pi}\frac{1}{|z_n-\zeta+re^{i\varphi}|^2}\,d\varphi\\
	&\leq O(\lambda^0)\,r\left(\inf_{\varphi\in[0,2\pi)}|z_n-\zeta+re^{i\varphi}|^2\right)^{-1},
  \end{align*}
  where $r$ denotes $(z'_n-z'_{n-1}).$ The term in brackets can be estimated as follows
  \begin{align*}
	|z_n-\zeta+re^{i\varphi}|^2=|w+re^{i\varphi}|^2&=(w'+r\cos\varphi)^2+(w''+r\sin\varphi)^2\\
	&\geq(|w'|-r)^2+(|w''|-r)^2\\
	&\geq2r(r-(|w'|+|w''|)).
  \end{align*}
  Since $k$ is in $[z'_n-\delta,z'_n+\delta]$ with $\delta$ being equal to
  $\frac{1}{2}(z'_n-z'_{n-1}),$ the real part of $\zeta$ will lie in the very
  same interval, which implies
  \begin{align*}
	|w+re^{i\varphi}|^2\geq r(z'_n-z'_{n-1}-2|w''|)\geq r(r-2|z''_n|)=r\,O(\lambda^{-1/2}).
  \end{align*}
  Thus, we can conclude that
  \begin{equation*}
	\|\chi_aR_0(k,\cdot)\|_\infty\leq O(\lambda^{1/2})|k-z_n|\qquad\forall\,k\in[z'_n-\delta,z'_n+\delta],
  \end{equation*}
  provided $\lambda$ is big enough. So it remains to show that
  the domain of $\zeta$ is contained in the interior of the circle defined
  by~\eqref{circle}. This is the case if
  \begin{equation*}
	|k-z_n|^2<(z'_n-z'_{n-1})^2\qquad\forall\,k\in[z'_n-\delta,z'_n+\delta].
  \end{equation*}
  But on this interval
  \begin{equation*}
	|k-z_n|^2\leq\delta^2+z''^2_n=\frac{1}{4}r^2+z''^2_n.
  \end{equation*}
  So for fixed $n,$ the variable $\zeta$ will lie in the interior of the
  circle~\eqref{circle}, if
  \begin{equation*}
	z''^2_n<\frac{3}{4}r^2=O(\lambda^{-1}).
  \end{equation*}
  For big enough $\lambda$ this condition is full filled, since $z''^2_n$ is of
  $O(\lambda^{-2}).$ Notice that $r$ is smaller than~$|z_n-z_{n-2}|.$ Hence, this
  condition will also ensure that the $k$ interval of interest, will lie within
  the annulus on which the Laurent expansion of $\phi$ converges.
\end{proof}


%% file: module33.tex




\chapter{Exponential Decay}
\label{ch:Results}

The purpose of this chapter is to show that the truncated Gamow functions
$\chi_aG_n$ decay exponentially within a certain time regime on the support of
$V_\lambda.$ Thereby, the last step of the heuristic discussion given in the
introduction, which is
\begin{equation}\label{eq:Approx_Time_Evolution}
  e^{-iHt}\psi_0(x)\sim\int\mathcal F\psi_0(k)\eta(k)G(x)e^{-i\frac{k^2}{2}t}\,dk\sim G(x)\,e^{-\frac{\Gamma}{2}\,t},
\end{equation}
will be turned into a precise statement. We will discover that the error which
is produced by this approximation decreases with increasing $\lambda.$ And after
we have determined the regime in which the exponential law is valid, we will
conclude by extending these results to initial wave functions which lie in
$\mathcal C^2([-a,a])$ and vanish at $\pm a.$

\section{Truncated Gamow Functions}\label{sec:Truncated}

The exponential decay of the truncated Gamow functions $\chi_aG_n$ within a
certain time regime follows from the exponential decay of any other initial
wave function lying in their neighborhood~(with respect to the $L^2$-norm). This
is due to the unitarity of $e^{-iH_\lambda t},$ which implies that an
$L^2$-neighbor~$\psi_n$ of the truncated Gamow function $\chi_aG_n$ satisfies
\begin{equation*}
  \|e^{-iH_\lambda t}\chi_aG_n-e^{-iH_\lambda t}\psi_n\|_2=\|\chi_aG_n-\psi_n\|_2\ll1.
\end{equation*}
So for now it will make no difference if we choose the bound states of
$H_\lambda$ restricted\linebreak[4]to $L_D=\{\psi\in L^2([-a,a])|\psi(a)=\psi(-a)=0\}$
in place of $\chi_aG_n.$ But we will benefit from this choice when the results
derived for these specific initial wave functions will be extended to generic
ones. Apart from that, we restrict ourselves to symmetric bound states for
convenience, and in order to see that they lie in the vicinity of~$\chi_aG_n,$
notice that the symmetric bound states are given by
\begin{equation*}
  \psi_n=\cos(\tilde z'_{2n-1}\,\cdot)\chi_a.
\end{equation*}
Therefore, they differ from the symmetric truncated Gamow functions only by the
appearance of the real part~$\tilde z'_{2n-1}$ instead of the entire
resonance~$\tilde z_{2n-1}.$ Given that the imaginary part of the resonance is
of~$O(\lambda^{-1/2})$ we only need to choose $\lambda$ big enough to ensure
that the distance between the symmetric bound states and the symmetric truncated
Gamow functions is small enough with respect to the $L^2$-norm.

In order to show that the time evolved symmetric bound states decay
exponentially within a certain time regime, we need to determine
$e^{-iH_\lambda t}\psi_n.$ In this regard, we will exploit that for big~$\lambda$ the
major part of the~$L^2$-mass of the energy distribution $\mathcal F\psi_n(k)$ is
concentrated in a small interval about~$z'_{2n-1}.$ This becomes evident from
\begin{equation}\label{eq:Psi_Hat}
  \mathcal F\psi_n(k)=\hat\psi_n(k)=\frac{c_n\,e^{ika}}{\bar F(k)}\,\frac{\cos \tilde ka}{z^{'2}_{2n-1}-k^2}\,.
\end{equation}
Its second factor resembles a $\mathrm{sinc}$-function which is peaked about
$z'_{2n-1}$ and has roots at $z'_{2m-1}$ for all $m\neq n.$ Due to the fact that
$|\bar F(z'_{2m})|^{-1}$ is of $O(\lambda^0)$ and $|\bar F(z'_{2m-1})|^{-1}$ is
of~$O(\lambda^{1/2})$ for all integer~$m,$ this causes~$|\hat\psi_n|$ to be
strongly peaked in a neighborhood of~$z'_{2n-1}$ while being~of~$O(\lambda^0)$
everywhere else. And since the distance between neighboring roots of the energy
distribution is of~$O(\lambda^{-1/2}),$~we conclude that the major part of
the~$L^2$-mass of~$\hat\psi_n$ is concentrated in the interval
\begin{equation}\label{eq:Interval}
  [z'_{2n-1}-\delta,z'_{2n-1}+\delta]\quad\text{with}\quad\delta=\frac{1}{2}(z'_{2n-1}-z'_{2n-2})=O(\lambda^{-1/2}).
\end{equation}

This property of the energy distribution $\hat\psi_n$ is the reason why the time
evolution of the symmetric bound states is governed by the generalized
eigenfunctions $\phi(k,\cdot)$ that belong to the interval~\eqref{eq:Interval}.
So this property is the reason for~\eqref{eq:Approx_Time_Evolution} to be a good
approximation. However, for the rigorous justification
of~\eqref{eq:Approx_Time_Evolution} we need to estimate the error that is
produced when~$\phi(k,\cdot)$ is replaced by~$\eta_n(k)G_n,$ which is its
leading order contribution on the interval~\eqref{eq:Interval}. Therefore, we
will consider
\begin{align*}
  \chi_ae^{-iH_\lambda t}\psi_n&=\chi_a\int_0^\infty\hat\psi_n(k)\eta_n(k)G_n(\cdot)e^{-i\frac{k^2}{2}t}\,dk\\
  &\quad+\chi_a\int_0^\infty\hat\psi_n(k)(\phi(k,\cdot)-\eta_n(k)G_n(\cdot))e^{-i\frac{k^2}{2}t}\,dk
\end{align*}
and prove that the $L^2$-norm of the error integral
\begin{equation*}
  E_n(\cdot,t)=\chi_a\int_0^\infty\hat\psi_n(k)(\phi(k,\cdot)-\eta_n(k)G_n(\cdot))e^{-i\frac{k^2}{2}t}\,dk
\end{equation*}
is small, while we will see that its main contribution
\begin{equation*}
  \Psi_n(\cdot,t)=\chi_a\int_0^\infty\hat\psi_n(k)\eta_n(k)G_n(\cdot)e^{-i\frac{k^2}{2}t}\,dk
\end{equation*}
decays exponentially within a certain time regime.

\section{Truncated Gamow Functions - Main Contribution}
\label{sec:Main_Contribution}

In the current section it will be shown that the main contribution $\Psi_n$ to
$\chi_ae^{-iH_\lambda t}\psi_n$ decays exponentially within a certain time regime and
moreover this regime will be determined.

\begin{figure}[t]
  \centering
  \includegraphics[scale=0.6]{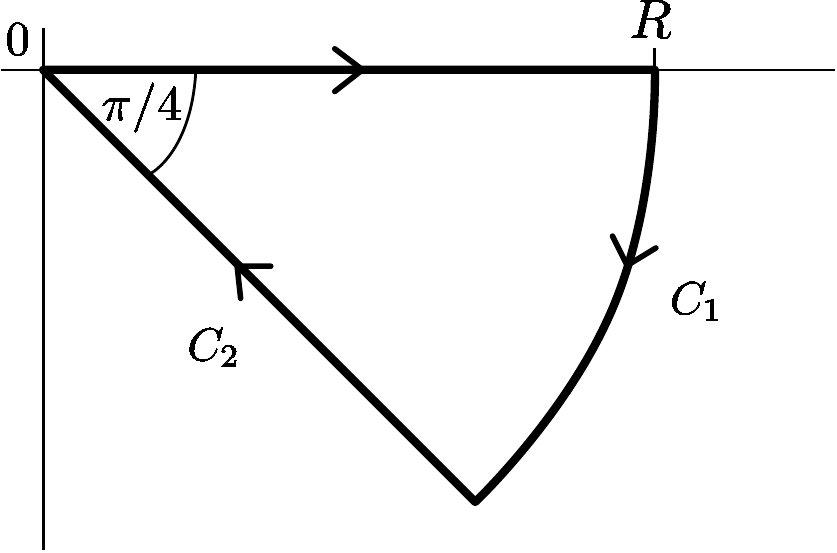}
  \caption{Illustration of the integration contour $C$ in the complex $k$-plane,
	where $R$ is understood to tend to $\infty.$}
  \label{fig:Contour}
\end{figure}

In order to prove that there is a time interval in which 
\begin{equation*}
  \Psi_n(\cdot,t)=\chi_aG_n\int_0^\infty\hat\psi_n(k)\,e^{-ika}\,\frac{a_{-1}}{k-z_{2n-1}}e^{-i\frac{k^2}{2}t}\,dk
\end{equation*}
decays exponentially, we will use the calculus of residues. Due to the term
$\exp(-i\frac{k^2}{2}t),$ the integration contour needs to run through the lower complex
$k$-plane in order to get converging integrals. In particular, we will choose the
contour illustrated in Figure~\ref{fig:Contour} with~$R$ tending to~$\infty,$
which results in an exponentially decaying contribution to~$\Psi_n$ caused by
the pole of~$\eta_n.$ Therefore, our main task will be to determine the
contribution from the rest of the integration contour and to show that it can be
neglected during a certain time regime.

\begin{lemm}\label{lem:Main_Contribution}
  Let $n$ be a fixed odd integer satisfying $1\leq n< n_\lambda.$ Then for
  all $t>0,$ we have
  \begin{equation}\label{eq:Main_Contribution}
	\Psi_n(\cdot,t)=c\,\chi_aG_ne^{-i\frac{z^2_n}{2}t}+\chi_aG_nr(t),
  \end{equation}
  where the non-zero constant $c$ is of $O(\lambda^0)$ and
  \begin{equation*}
	|r(t)|\leq O(\lambda^{-2})\,t^{-1/2}.
  \end{equation*}
\end{lemm}
\begin{proof}
  By the residue theorem
  \begin{equation*}
	\oint_C\hat\psi_n(z)\eta_n(z)e^{-i\frac{z^2}{2}t}\,dz=-2\pi
	i\,\res{z_n}{\hat\psi_n\eta_ne^{-i\frac{t}{2}(\cdot)^2}},
  \end{equation*}
  where $C$ is the contour illustrated in Figure~\ref{fig:Contour}. From this
  figure it becomes clear that the contour integral breaks up into
  \begin{equation*}
	\int_0^\infty\hat\psi_n(k)\eta_n(k)e^{-i\frac{k^2}{2}t}\,dk+\int_{C_1+C_2}\hat\psi_n(z)\eta_n(z)e^{-i\frac{z^2}{2}t}\,dz.
  \end{equation*}

  We will show first that the arc-like part $C_1$ of the integration contour
  gives no contribution. In this respect, notice that
  \begin{equation*}
	|\int_{C_1}|\leq\lim_{R\rightarrow\infty}R\int_0^{\pi/4}|\hat\psi_n(Re^{-i\varphi})\eta_n(Re^{-i\varphi})|\,e^{-\frac{t}{2}R^2\sin2\varphi}\,d\varphi.
  \end{equation*}
  As $\lim_{R\rightarrow\infty}|\hat\psi_n(Re^{-i\varphi})e^{-iaRe^{-i\varphi}}|=0$
  for all~$\varphi$~in~$[0,\pi/4],$ which becomes evident by a straightforward
  calculation, we only need to choose $R$ big enough to ensure the following
  estimates
  \begin{align*}
	|\int_{C_1}|&\leq\lim_{R\rightarrow\infty}cR\int_0^{\pi/4}((R\cos\varphi-z'_n)^2+(R\sin\varphi+z''_n)^2)^{-1/2}\,e^{-\frac{t}{2}R^2\sin2\varphi}\,d\varphi\\
	&\leq\lim_{R\rightarrow\infty}cR\int_0^{\pi/4}|R\cos\varphi-\frac{R}{2}\cos\varphi|^{-1}\,e^{-\frac{t}{2}R^2\frac{2}{\pi}\varphi}\,d\varphi\\
	&\leq\lim_{R\rightarrow\infty}c'\int_0^{\pi/4}e^{-\frac{t}{2}R^2\frac{2}{\pi}\varphi}\,d\varphi=\lim_{R\rightarrow\infty}c''\left(\frac{1-e^{-t\frac{R^2}{4}}}{tR^2}\right)=0.
  \end{align*}

  Let us estimate the modulus of the contribution from $C_2$ next. Using the
  explicit form of~$\hat\psi_n$ and~$\eta_n,$ the following estimates are
  immediately evident
  \begin{align*}
	|\int_{C_2}|&\leq \int_0^\infty\frac{|c_n|}{|1+i\frac{\tilde z}{z}\tan a\tilde z|}\,\frac{1}{|z^{'2}_n+ir^2|}\,\frac{|a_{-1}|}{|z-z_n|}\,e^{-\frac{t}{2}r^2}\,dr\\
	&\leq \frac{|c_n\,a_{-1}|}{z'^2_n} \int_0^\infty\big|1+i\frac{\tilde z}{z}\tan a\tilde z\big|^{-1}\,|z-z_n|^{-1}\,e^{-\frac{t}{2}r^2}\,dr.
  \end{align*}
  To handle the first factor of the remaining integrand, we will show that
  $\Re(i\frac{\tilde z}{z}\tan a\tilde z)$ is positive, since this implies
  the following inequality
  \begin{equation*}
  	\big|1+i\frac{\tilde z}{z}\tan a\tilde z\big|^{-1}<\big|\frac{\tilde z}{z}\tan a\tilde z\big|^{-1}.
  \end{equation*}
  In order to see this, notice that
  \begin{align*}
	&\arg\left(i\frac{\tilde z}{z}\tan a\tilde z\right)=\frac{3}{4}\pi+\arg\left(i\sqrt{2\lambda+ir^2}\right)+\arg(\tan a\tilde z)\quad\text{and}\\
	&\tan a\tilde z=\frac{\sin 2a\tilde z'+i\sinh 2a\tilde z''}{\cos 2a\tilde z'+\cosh 2a\tilde z''}.
  \end{align*}
  Therefore, $\tilde z''\geq\sqrt{2\lambda}$ and this implies that
  \begin{equation*}
	\arg(\tan a\tilde
	z)\in\left(\frac{\pi}{4},\frac{3}{4}\pi\right)\quad\text{as well as}\quad|\tan a\tilde z|>O(\lambda^0)\neq0.
  \end{equation*}
  But then $\arg(i\frac{\tilde z}{z}\tan a\tilde z)$ remains within the interval
  $(3\pi/2,2\pi+\pi/4),$ which justifies
  \begin{equation*}
	|\int_{C_2}|<c\frac{|c_n\,a_{-1}|}{z'^2_n}\int_0^\infty |re^{-i\pi/4}-z_n|^{-1}\,e^{-\frac{t}{2}r^2}\,dr
  \end{equation*}
  provided $\lambda$ is big enough. By determining the maximum of the first term
  in the remaining integrand and using Theorem~\ref{thm:Laurent_Coefficients}, we
  can thereby conclude that
  \begin{equation*}
	|\int_{C_2}|<c\frac{|c_n\,a_{-1}|}{z'^2_n(z'_n-|z''_n|)}\int_0^\infty e^{-\frac{t}{2}r^2}\,dr\leq O(\lambda^{-2})\,t^{-1/2}.
  \end{equation*}

  So it remains to determine the residue and its order with respect
  to~$\lambda.$ First of all, it is not difficult to see that
  \begin{equation*}
	\res{z_n}{\hat\psi_n\eta_ne^{-i\frac{t}{2}(\cdot)^2}}=a_{-1}\hat\psi_n(z_n)e^{-iaz_n}e^{-i\frac{z_n^2}{2}t}=c\,e^{-i\frac{z_n^2}{2}t}.
  \end{equation*}
  From the previous section it is known that $|\hat\psi_n(z'_n)|$ is
  of~$O(\lambda^{1/2}).$ Since~$z''_n$ is of~$O(\lambda^{-1}),$ the continuity
  of~$\hat\psi_n$ in the region between $z_n$ and the real axis
  requires~$|\hat\psi_n(z_n)|$ to be of~$O(\lambda^{1/2}).$ But according to
  Theorem~\ref{thm:Laurent_Coefficients}, we
  have~$|a_{-1}|\leq O(\lambda^{-1/2}).$ So we conclude that~$c$ is
  of~$O(\lambda^0).$
\end{proof}

To determine the regime in which $\Psi_n$ decays exponentially consider the
ratio of the two summands that contribute to
$\Psi_n$~(equation~\eqref{eq:Main_Contribution}), that is
\begin{equation*}
  \frac{|r(t)|}{\left|c\,e^{-i\frac{z_n^2}{2}t}\right|}\leq O(\lambda^{-5/4})\,e^{\Gamma_nt}\,(\Gamma_nt)^{-1/2}.
\end{equation*}
According to Lemma~\ref{lem:Main_Contribution} exponential decay prevails, if
this quotient is much smaller than one. Clearly, this can not be the case for
$t\ll 1$ nor for $t\gg 1.$ But this is exactly what we expected, since it was
shown in Section~\ref{sec:Barry} that the survival probability
$|\langle\psi_0,e^{-iHt}\psi_0\rangle|^2$ can not decay exponentially in these time
regimes as long as~$\psi_0$ is a non-zero element of~$\mathcal D(H).$ And the
initial wave functions~$\psi_n$ are elements of~$\mathcal D(H_\lambda),$ because
their weak derivatives~$\psi'_n$ and~$\psi''_n$ exist and have
finite~$L^2$-norm. However, on intermediate time scales the above-mentioned
quotient is small for periods of several lifetimes provided $\lambda\gg1.$ This
becomes immediately evident when~$t$ is such that it satisfies
\begin{equation*}
  1\leq\Gamma_nt\leq N
\end{equation*}
with an integer $N>1$ that depends on $\lambda.$

\section{Truncated Gamow Functions - Error}

To finish the proof that~$e^{-iH_\lambda t}\psi_n$ decays exponentially in a certain time
interval on the support of~$V_\lambda,$ it remains to estimate the $L^2$-norm of
the error integral
\begin{equation*}
  E_n(\cdot,t)=\chi_a\int_0^\infty\hat\psi_n(k)(\phi(k,\cdot)-\eta_n(k)G_n(\cdot))e^{-i\frac{k^2}{2}t}\,dk.
\end{equation*}
We will see that this norm is of $O(\lambda^{-1/4}).$ Therefore, the main
contribution determined in the previous section will approximate
$\chi_ae^{-iH_\lambda t}\psi_n$ better the bigger $\lambda$ is chosen. This is in accord with
our intuition that an initially localized wave function remains longer within
the support of $V_\lambda$ the higher the potential plateau is.

Let us illustrate briefly why we should expect $E_n(\cdot,t)$ to have small
$L^2$-norm before it will be proven rigorously. It was explained in
Section~\ref{ch:Results}.\ref{sec:Truncated} that the energy distribution
$\hat\psi_n$ is strongly localized within the interval $\mathcal I$ given by
$[z'_{2n-1}-\delta,z'_{2n-1}+\delta].$ By exploiting this fact we will be able
to show that the contribution from $\mathbb R^+\setminus\mathcal I$ to $E_n$ has
small $L^2$-norm. To estimate the contribution from $\mathcal I$ itself, we will
use the fact that $\eta_nG_n$ is the principal part of the Laurent expansion of
$\phi.$ But since this part of the proof is a bit more involved, we will briefly
outline the approach. Using Lemma~\ref{lem:Laurent} and
Theorem~\ref{thm:Laurent_Coefficients}, the contribution from $\mathcal I$ can
be expressed as
\begin{equation*}
  \chi_aa_0(\cdot)\int_{\mathcal I}\hat\psi_n(k)e^{-ika}e^{-i\frac{k^2}{2}t}\,dk
  +\chi_a\int_{\mathcal I}\hat\psi_n(k)e^{-ika}R_0(k,\cdot)e^{-i\frac{k^2}{2}t}\,dk.
\end{equation*}
The details on the remainder $R_0$ proven in
Theorem~\ref{thm:Laurent_Coefficients} are sufficient to handle the second
integral in a straightforward manner. The involved part is the first integral.
In this respect, notice that~$\hat\psi_n$ contains~$\bar F^{-1}$ whose
meromorphic continuation has poles at~$\bar z_n$ instead of~$z_n.$
Therefore,~$\bar F^{-1}$ has a Laurent expansion which is completely analogous
to the one of~$F^{-1},$ namely
\begin{equation*}
  \bar F^{-1}(k)=\frac{a_{-1}}{k-\bar z_n}+R(k).
\end{equation*}
From this expression it becomes clear that the phase of the principal
part of $\hat\psi_n$ oscillates rapidly on the interval $\mathcal I.$ Therefore,
we will use a stationary phase argument to handle the major contribution to the
first integral from above.


\begin{lemm}\label{lem:One-Resonance-Approx}
  Let $n$ be a fixed odd integer satisfying $1\leq n< n_\lambda.$
  Then for all $t>0,$ we have
  \begin{equation*}
	\|E_n(\cdot,t)\|_2\leq O(\lambda^{-1/4}).
  \end{equation*}
\end{lemm}

\begin{proof}
  For brevity let $\mathcal I$ denote $[z'_n-\delta,z'_n+\delta],$ where
  $\delta$ can be chosen as $\frac{1}{2}(z'_{n}-z'_{n-1})$ due to
  Theorem~\ref{thm:Laurent_Coefficients}. Then,
  \begin{align*}
	\|E_n(\cdot,t)\|_2\leq
	&\left\|\chi_a\int_0^\infty(1-\chi_{_\mathcal I})(k)\hat\psi_n(k)(\phi(k,\cdot)-\eta_n(k)G_n(\cdot))e^{-i\frac{k^2}{2}t}\,dk\right\|_2\\
	&+\left\|\chi_a\int_\mathcal I\hat\psi_n(k)(\phi(k,\cdot)-\eta_n(k)G_n(\cdot))e^{-i\frac{k^2}{2}t}\,dk\right\|_2\\
	&=:A+B.
  \end{align*}

  First consider $A.$ The following estimate is immediately evident
  \begin{align*}
	A&\leq\,\left\|\int_0^\infty(1-\chi_{_\mathcal I})(k)\hat\psi_n(k)\phi(k,\cdot)e^{-i\frac{k^2}{2}t}\,dk\right\|_2\\
	&\qquad+\|\chi_aG_n\|_2\,\int_0^\infty(1-\chi_{_\mathcal I})(k)|\hat\psi_n(k)\eta_n(k)|\,dk\\
	&\quad=:A_1+A_2,
  \end{align*}
  where $\chi_a$ was dropped in the first summand. From the spectral theorem, we
  deduce
  \begin{align*}
	A_1^2&=\left\|e^{-iH_\lambda t}\int_0^\infty(1-\chi_{_\mathcal I})(k)\hat\psi_n(k)\phi(k,\cdot)\,dk\right\|_2^2
	=\left\|e^{-iH_\lambda t}\mathcal F^{-1}(1-\chi_{_\mathcal I})\hat\psi_n\right\|_2^2.
  \end{align*}
  Using the unitarity of $\mathcal F$ and $e^{-iH_\lambda t},$ this reduces to
  \begin{align*}
	A_1^2=\int_0^\infty(1-\chi_{_\mathcal I})(k)|\hat\psi_n(k)|^2\,dk
	\leq\int_0^\infty(1-\chi_{_\mathcal	I})(k)\frac{|c_n|^2}{(z'^2_n-k^2)^2}\,dk\leq O(\lambda^{-1/2}),
  \end{align*}
  where the last step follows from a straightforward calculation. Regarding
  $A_2,$ notice that
  \begin{equation*}
  	\|\chi_aG_n\|_\infty=\|\chi_a\cos(\tilde z_n\cdot)\|_\infty\leq O(\lambda^0),
  \end{equation*}
  which follows from the $\lambda$-dependence of $\tilde z_n$ given in
  Lemma~\ref{lem:Resonances}. Therefore, we find the following estimates
  \begin{align*}
	A_2&\leq c\sup_{k\in\mathbb R^+\setminus\mathcal I}\frac{|a_{-1}|}{|k-z_n|}\int_0^\infty(1-\chi_{_\mathcal I})(k)\frac{|c_n|}{|z'^2_n-k^2|}\,dk\\
	&\leq c'\int_0^\infty(1-\chi_{_\mathcal I})(k)\frac{|c_n|}{|z'^2_n-k^2|}\,dk\\
	&\leq O(\lambda^{-1/2}),
  \end{align*}
  where the last step follows again from a straightforward calculation. And this
  directly implies that
  \begin{equation*}
	A\leq O(\lambda^{-1/4})\quad\forall\,t>0.
  \end{equation*}

  So it remains to estimate $B.$ From Lemma~\ref{lem:Laurent} and
  Theorem~\ref{thm:Laurent_Coefficients}, it is known that
  \begin{align*}
	&\phi(k,\cdot)-\eta_n(k)G_n(\cdot)=e^{-ika}\left(a_0(\cdot)+R_0(k,\cdot)\right)\qquad\forall\,k\in\mathcal I\\
	&\text{with}\quad\|\chi_aa_0\|_\infty\leq O(\lambda^0)\quad\text{and}\quad\|\chi_aR_0(k,\cdot)\|_\infty\leq O(\lambda^{1/2})|k-z_n|.
  \end{align*}
  Thus, using the fact that $x$ takes only values within a compact set we find
  \begin{align*}
	B&\leq O(\lambda^0)\left|\int_{\mathcal I}\hat\psi_n(k)e^{-ika}e^{-i\frac{k^2}{2}t}\,dk\right|
	+O(\lambda^{1/2})\,\delta\,\sup_{k\in\mathcal I}|\hat\psi_n(k)(k-z_n)|=:B_1+B_2.
  \end{align*}
  If we define
  \begin{equation*}
	h_n(k)=\frac{\cos\tilde ka}{z'^2_n-k^2}
  \end{equation*}
  for brevity and apply the results of Theorem~\ref{thm:Laurent_Coefficients}
  to~$F^{-1},$ we find
  \begin{align*}
	\sup_{k\in\mathcal I}|\hat\psi_n(k)(k-z_n)|&=|c_n|\,\sup_{k\in\mathcal I}|\bar F^{-1}(k)(k-z_n)h_n(k)|\\
	&\leq c'_n\,\sup_{k\in\mathcal I}|F^{-1}(k)(k-z_n)|\\
	&\leq c'_n\,\sup_{k\in\mathcal I}(|a_{-1}|+|a_0(k-z_n)|+|R_0(k)(k-z_n)|)=O(\lambda^{-1/2}).
  \end{align*}
  And due to the fact that $\delta$ is of $O(\lambda^{-1/2}),$ we conclude
  \begin{equation*}
	B_2\leq O(\lambda^{-1/2}).
  \end{equation*}
  As was described above, we want to apply a stationary phase argument to
  estimate~$B_1.$ For this purpose we will use
  Theorem~\ref{thm:Laurent_Coefficients} once again in order to expand~$\bar
  F^{-1},$ which is part of $\hat\psi_n.$ This immediately implies
  \begin{align*}
	B_1&\leq O(\lambda^0)\left|\int_{\mathcal I}\frac{a_{-1}}{k-\bar z_n}\,h_n(k)e^{-i\frac{k^2}{2}t}\,dk\right|
	+O(\lambda^0)\,\delta\,\sup_{k\in\mathcal I}|(a_0+R_0(k))h_n(k)|\\
	&\leq O(\lambda^0)\left|\int_{\mathcal I}\tilde\eta_n(k)\,h_n(k)e^{-i\frac{k^2}{2}t}\,dk\right|+O(\lambda^{-1/2}),
  \end{align*}
  where $\tilde\eta_n$ was defined as $a_{-1}(k-\bar z_n).$ Since $h_n$ is real
  valued, it does not contribute to the phase of the remaining integrand.
  Therefore, we can use partial integration in order to get
  \begin{align}\label{eq:proof_error}
	|\int_{\mathcal I}|&=\left|\int_{\mathcal I}|\tilde\eta_n(k)|h_n(k)e^{iS(k)-i\frac{k^2}{2}t}\,dk\right|\nonumber\\
	&\leq\left|\left[\frac{|\tilde\eta_n(k)|h_n(k)}{\partial_kS(k)-kt}e^{iS(k)-i\frac{k^2}{2}t}\right]_{z'_n-\delta}^{z'_n+\delta}\right|
	+\int_{\mathcal	I}\left|\partial_k\frac{|\tilde\eta_n(k)|h_n(k)}{\partial_kS(k)-kt}\right|\,dk,
  \end{align}
  where~$S$ denotes the phase of~$\tilde\eta_n.$ Due to the fact
  that~$\tilde\eta_n$ is of~$O(\lambda^0)$ at the boundaries of integration, as
  well as
  \begin{equation*}
	\partial_kS(k)=\partial_k\arctan\frac{-z''_n}{k-z'_n}=\frac{z''_n}{(k-z'_n)^2+z''^2_n}<0,
  \end{equation*}
  we immediately see that the first summand in~\eqref{eq:proof_error} is bounded
  by a constant of~$O(\lambda^{-1/2})$ for all~$t>0.$
  
  The second term that appears in~\eqref{eq:proof_error} will be of the same
  order, if the supremum of its integrand is bounded by a constant of
  $O(\lambda^0).$ An application of the product rule yields the following upper
  bound for the integrand
  \begin{equation*}\label{eq:proof_error2}
	\left\|\chi_{_\mathcal I}\frac{h_n|\tilde\eta_n|'}{S'-(\cdot)t}\right\|_\infty
	+\left\|\chi_{_\mathcal I}\frac{|\tilde\eta_n|h'_n}{S'-(\cdot)t}\right\|_\infty
	+\left\|\chi_{_\mathcal	I}\frac{S''-t}{(S'-(\cdot)t)^2}|\tilde\eta_n|h_n\right\|_\infty.
  \end{equation*}
  Straightforward calculations justify the following estimates
  \begin{align*}
	&\left\|\chi_{_\mathcal I}|\tilde\eta_n|'\right\|_\infty=
	\left\|\chi_{_\mathcal I}\frac{|a_{-1}|(k-z'_n)}{((k-z'_n)^2+z''^2_n)^{3/2}}\right\|_\infty\leq O(\lambda^{1/2})\\
	\text{and }&\left\|\chi_{_\mathcal I}S''\right\|_\infty=
	\left\|\chi_{_\mathcal I}\frac{2z''_n(k-z'_n)}{((k-z'_n)^2+z''^2_n)^2}\right\|_\infty\leq O(\lambda^{1/2}).
  \end{align*}
  To get an upper bound for $\|\chi_{_\mathcal I}h'_n\|_\infty$ notice that $h_n$ is an element
  of the Sobolev space $H^2(\mathbb R).$ We can therefore apply Sobolev's
  inequality~(see~\cite[Chapter~8]{LiebLoss}) in order to find that
  \begin{align*}
	\|\partial_kh_n(k)\|^2_\infty\leq\frac{1}{2}\left(\|\partial_kh_n(k)\|_2^2+\|\partial^2_kh_n(k)\|_2^2\right)
	\leq\frac{1}{2}\left(\|x\psi_n(x)\|_2^2+\|x^2\psi_n(x)\|_2^2\right)\leq O(1),
  \end{align*}
  where we used the fact that $h_n$ is the usual Fourier transform of $\psi_n.$
  Putting all these inequalities together we find
  \begin{align*}
	\left\|\chi_{_\mathcal I}\left(\frac{|\tilde\eta_n|h_n}{S'-(\cdot)t}\right)'\right\|_\infty&\leq
	O(\lambda^{1/2})\left\|\frac{\chi_{_\mathcal I}}{S'-(\cdot)t}\right\|_\infty
	+O(\lambda)\left\|\frac{\chi_{_\mathcal I}}{(S'-(\cdot)t)^2}\right\|_\infty
	+O(\lambda^0)\left\|\chi_{_\mathcal I}\frac{|\tilde\eta_n|}{S'-(\cdot)t}\right\|_\infty\\
	&\leq O(1)\qquad\forall\,t>0.
  \end{align*}
  This shows that $B_1$ has to be bounded by a constant of $O(\lambda^{-1/2}),$
  which allows us to conclude that
  \begin{equation*}
	\|E_n(\cdot,t)\|_2\leq A+B\leq O(\lambda^{-1/4})\qquad\forall\,t>0.
  \end{equation*}
\end{proof}

\section{Generic Initial Wave Functions}

From a physical point of view it is not satisfactory, that only the specific
initial wave functions considered in the previous sections show exponential
decay in certain time intervals. Therefore, we will extend the results achieved
so far to generic~$\psi_0$ in this section. In this process we will understand
why truncated Gamow functions are distinguished initial wave functions. And
moreover, it will become clear why the bound states of~$H_\lambda$ restricted
to~$L_D$ are better adapted for the extension to generic initial wave
functions than the truncated Gamow functions.

But what does generic mean in the first place? The probability that the
meta-stable particle already decayed should initially still be zero. So we still
demand that $supp~\psi_0\subset[-a,a].$ For simplicity we will restrict
ourselves to symmetric wave functions again. And since we exploited the strong
localization of $\mathcal F\psi_0,$ the Riemann-Lebesgue Lemma suggests
that~$\psi_0$ should be sufficiently smooth. Therefore, under generic we shall
understand that
\begin{equation*}
  \psi_0\in\left\{f\in\mathcal C^k_0([-a,a])|f(x)=f(-x)\right\}=:\mathcal A_k,
\end{equation*}
where $k$ will be chosen big enough and $\mathcal C^k_0([-a,a])$ denotes the
space of $k$-times continuously differentiable functions on $[-a,a]$ which
vanish at $\pm a.$

The fact that the generalized eigenfunction $\phi(k,\cdot)$ could be replaced
in~\eqref{eq:Approx_Time_Evolution} by the principal part of its Laurent
expansion about $z_n$ without producing a huge error, was possible only since
the contribution $\|\chi_{\mathbb R^+\setminus[z'_n-\delta,z'_n+\delta]}\mathcal
F\psi_0\|_2$ from all the other resonances was small. So for the method from the
previous section to be successful, this one-resonance-approximation has to
remain valid for elements of $\mathcal A_k.$ However, it is not difficult to see
that this will not happen without any further restrictions on $\psi_0,$ even
when $k=\infty.$ Choose for example
\begin{equation*}
  \psi_0(x)=(\cos\tilde z'_1x+\cos\tilde z'_3x)\chi_a(x),
\end{equation*}
then this will be an element of $\mathcal A_k$ and we know from
Section~\ref{ch:Results}.\ref{sec:Truncated} that
\begin{equation*}
  \mathcal F\psi_0(k)=\frac{e^{ika}}{\bar F(k)}\left(\frac{c_1\cos\tilde ka}{z'^2_1-k^2}+\frac{c_3\cos\tilde ka}{z'^2_3-k^2}\right).
\end{equation*}
But this is strongly localized about $z'_1$ and $z'_3,$ which implies that
$\|\chi_{\mathbb R^+\setminus[z'_1-\delta,z'_1+\delta]}\mathcal F\psi_0\|_2$ is
not small. This shows that the initial wave functions 
\begin{equation*}
  \mathcal B_0=\left\{\cos(\tilde z'_n\,\cdot)\chi_a\right\}_{n\in\mathbb N}
\end{equation*}
considered in the previous sections are distinguished insofar as they allow
for a one-resonance-approximation. And since the truncated Gamow function
$\cos(\tilde z_n\,\cdot)\chi_a$ belongs to the neighborhood of
$\cos(\tilde z'_n\,\cdot)\chi_a$ (measured in the $L^2$-norm) it is
distinguished for the same reason. But for generic initial wave functions there
will be contributions from more than one resonance. And in order to handle them,
we will follow the hint that is actually given by the above counterexample.

As already noted in Section~\ref{ch:Results}.\ref{sec:Truncated}, the initial
wave functions $\mathcal B_0$ are a subset of
\begin{equation*}
  \mathcal B=\left\{\sin\frac{n\pi(x+a)}{2a}\right\}_{n\in\mathbb N},
\end{equation*}	
which are the eigenfunctions of $H_0$ with domain $\mathcal C^2_0([-a,a]).$ From
the theory of Fourier series~(see~\cite[Example~1.2.3]{Davies}) it is well-known that
$\mathcal B$ is a complete orthonormal set in~$L^2([-a,a]).$ One
consequence of this is the essential self-adjointness of $H_0$ on its domain.
But more importantly we can conclude that $\mathcal B_0$ is an orthonormal basis
in the set of general initial wave functions $\mathcal A_k.$ The counterexample
given above is therefore just one particular element of $\mathcal A_k$ expressed
as a Fourier series, which as we now know can be done with any other element as
well.

Thus, we can now determine the time evolution of our general class of initial
wave functions through a term by term application of the
one-resonance-approximation, which yields
\begin{align}\label{eq:Multi_Resonance}
  &e^{-iH_\lambda t}\psi_0(x)=\sum_{n=1}^\infty a_nC_n(t)G_n(x)+R(x,t)\quad\text{for all}\quad x\in[-a,a],\nonumber\\
  &\text{where}\quad C_n(t)=\int_0^\infty\mathcal F\psi_n(k)\eta_n(k)e^{-i\frac{k^2}{2}t}\,dk
  \quad\text{with}\quad\psi_n\in\mathcal B_0.
\end{align}
The exponential decay within a certain time regime then immediately follows from
Lemma~\ref{lem:Main_Contribution}. And to avoid questions of summability we will
exploit the smoothness of the elements in~$\mathcal A_k.$ This property together
with the Riemann-Lebesgue Lemma ensure that the modulus of the Fourier
coefficients $a_n$ is as small as we please if only $n$ is big enough.
Therefore, it will be sufficient to apply the one-resonance-approximation to the
first few summands, which represent the major contribution to $\psi_0.$

Before we will use the preceding considerations to prove the next Lemma, notice
that contrary to~$\mathcal B_0$ it is not evident that the set of truncated
Gamow functions~$\left\{\chi_aG_n\right\}_{n\in\mathbb N}$ is a basis
in~$\mathcal A_k.$ Therefore, it is not clear which set of initial wave functions
actually is covered by finite linear combinations of them.

\begin{lemm}\label{lem:General}
  Let $\psi_0$ be an element of $\mathcal A_1$ and let $C_n(t)$ be defined as in
  equation~\eqref{eq:Multi_Resonance}. If $\lambda$ is big enough, there is an
  integer $1\leq N< n_\lambda$ such that for every $x\in[-a,a]$
  \begin{equation*}
	e^{-iH_\lambda t}\psi_0(x)=\sum_{n=1}^Na_nC_n(t)G_n(x)+R(x,t)
  \end{equation*}
  with $\|\chi_aR(\cdot,t)\|_2\leq O(\lambda^{-1/4})$ for all $t>0.$
\end{lemm}

\begin{proof}
  Let $\psi_n$ with $n\geq1$ denote the elements of $\mathcal B_0.$ Following
  the considerations from above these elements build a complete orthonormal set
  in $\mathcal A_k.$ Hence, for every $\psi_0$ in $\mathcal A_k$ we have
  \begin{align*}
	e^{-iH_\lambda t}\psi_0=e^{-iH_\lambda t}\sum_{n=1}^\infty\left<\psi_0,\psi_n\right>\psi_n&=
	e^{-iH_\lambda t}\sum_{n=1}^N a_n\psi_n+e^{-iH_\lambda t}\sum_{n=N+1}^\infty a_n\psi_n\\
	&= e^{-iH_\lambda t}\tilde\psi_0+R(\cdot,t),
  \end{align*}
  where the integer $N\geq1$ will be specified later.
  
  The Fourier coefficients $a_n$ are independent of $\lambda,$ since neither
  $\psi_0$ nor any of the $\psi_n$ depends on this parameter. Therefore, we can
  use the Riemann-Lebesgue Lemma together with the differentiability of
  $\psi_0$, in order to see that for any $m\leq k$ there is
  a~$\lambda$-independent constant~$C_m>0$ such that
  \begin{equation*}
	|a_n|\leq\frac{C_m}{n^m}\qquad\forall\,n\geq1.
  \end{equation*}
  This inequality together with the unitarity of $e^{-iH_\lambda t}$ imply
  \begin{equation*}
 	\|R(\cdot,t)\|_2^2=\bigg\|\sum_{n=N+1}^\infty a_n\psi_n\bigg\|_2^2
	=\sum_{n=N+1}^\infty |a_n|^2\leq C_k^2\sum_{n=N+1}^\infty n^{-2k},
  \end{equation*}
  where Plancherel's identity was used in the second step. For this sum to
  converge, it is sufficient to choose $k=1.$ In this case we find the
  following upper bound
  \begin{equation*}
	\|R(\cdot,t)\|_2^2\leq C_1^2\sum_{n=N+1}^\infty\int_{n-1}^n m^{-2}\,dm=\frac{C_1^2}{N}.
  \end{equation*}
  By Lemma~\ref{lem:One-Resonance-Approx} we can therefore find a
  $1\leq N< n_\lambda$ such that the one-resonance-approximation applies to
  every summand of $e^{-iH_\lambda t}\tilde\psi_0,$ while
  $\|R(\cdot,t)\|_2\leq O(\lambda^{-1/4})$ for all $t>0.$
  
  And using Lemma~\ref{lem:One-Resonance-Approx} once again, we conclude that for
  these $N$ and every $x\in[-a,a]$
  \begin{equation*}
	e^{-iH_\lambda t}\psi_0=\sum_{n=1}^N a_n\tilde\psi_n(t)+R'(\cdot,t),
  \end{equation*}
  where $\tilde\psi_n(t)$ denotes the main contribution from the
  one-resonance-approximation applied to~$e^{-iH_\lambda t}\psi_n$ that is
  \begin{equation*}
    \tilde\psi_n(t)=G_n\,\int_0^\infty\mathcal F\psi_n(k)\eta_n(k)e^{-i\frac{k^2}{2}t}\,dk
  \end{equation*}
  and $\|R'(\cdot,t)\|_2\leq O(\lambda^{-1/4})$ for all $t>0.$ 
\end{proof}


%% file: module4.tex



\chapter{Outlook}

In the previous chapters we studied the physical significance of Gamow functions
by analyzing the Schrödinger evolution of initially localized wave
functions on the support of the potential~$V_\lambda.$ The piece missing to
complete picture is therefore the evolution of these initial wave functions
outside of the support. In this regard, Skibsted's
results~\cite{Skibsted86} suggest that~$|e^{-iH_\lambda t}\psi_0(x)|$ increases
exponentially in $x$ up to a radius $R(t)$ which moves outward, since he proved
\begin{equation*}
  e^{-iH_\lambda t}\chi_aG_n\,\,\overset{L^2}{\sim}\,\,e^{-itz_n^2/2}\chi_{R(t)}G_n\quad\text{with}
  \quad R(t)=a+\mathrm{Re}(z_n)\,t.
\end{equation*}
On the one hand this is an astonishing assertion, but on the other hand large
radii $R(t)$ correspond to early escape. Thus, the exponential decay in time on
the support of the potential should cause an exponential increase in $x,$
otherwise the continuity equation would be violated. However, Skibsted's result
was proven in the $L^2$-norm, which raises the question whether the pointwise
exponential increase of the wave function is actually true. Proving this with
the help of an expansion in generalized eigenfunctions seems to be more delicate
than proving the exponential decay within a certain time regime on the support
of $V_\lambda.$ This is due to the fact that the main contribution to the
integral
\begin{equation*}
  e^{-iH_\lambda t}\psi_0(x)=\int\mathcal F\psi_0(k)\phi(k,x)e^{-i\frac{k^2}{2}t}\,dk,
\end{equation*}
which is its residue, is of~$O(\lambda^{-1/2})$ for~$x>a.$ Since the
lifetime~$\Gamma^{-1}_n$ increases with increasing~$\lambda,$ this is sensible
from a physical point of view. But from a mathematical viewpoint the problem
becomes more demanding, because this is even a lower order than some of the
error integrals obtained when the time evolution on the support of the potential
was studied.

The question whether the Gamow functions have physical relevance, which was the
subject this thesis, is part of the larger endeavor to study the Time-Energy
Uncertainty Relation. This question appeared due to the connection between
energy width and lifetime of meta-stable states. And although, the
$\alpha$-decay is a well-established application of the Time-Energy Uncertainty
Relation, this inequality is surrounded by many mathematical questions. One of
them is its rigorous derivation, which is problematic given that no self-adjoint
time operator exists. However, this as well as several other questions
concerning the Time-Energy Uncertainty Relation remain as a challenge for future
work.
